\documentclass[journal]{IEEEtran}

\usepackage{hyperref}
\usepackage{amsthm}
\usepackage{stfloats} 
\usepackage{array,cite}
\usepackage{pgfplots}
\usepgfplotslibrary{units}
\usetikzlibrary{patterns}
\usepackage{multirow} 
\usepackage{stmaryrd}
\usepackage{tikz}
\usepackage{booktabs}
\usetikzlibrary{arrows,calc,positioning}
\usepackage{footnote}
\usepackage{amsmath,amssymb}
\usepackage{makecell}
\usepackage{float}
\usepackage[ruled,noend]{algorithm2e}
\usepackage{algorithmic}
\pgfplotsset{compat=1.16}

\definecolor{IvoryBlack}{rgb}{0.1600, 0.1400, 0.1300}
\definecolor{GreenDark}{rgb}{0,0.3922,0}
\definecolor{RedLight}{rgb}{0.8922,0,0}
\definecolor{BlueLight}{rgb}{0, 0.3, 0.6039}
\definecolor{OrangeLight}{rgb}{0.9500, 0.6600, 0.4400}
\definecolor{VioletLight}{rgb}{0.5765, 0.4392, 0.8588}
 
\DeclareMathOperator\arctanh{arctanh}

\newtheorem{proposition}{Proposition}
\newtheorem{observation}{Observation}
\newtheorem{examplex}{Example}
\newenvironment{example}
  {\pushQED{\qed}\examplex}
  {\popQED\endexamplex}

\DeclareMathOperator*{\argmax}{arg\,max}
\DeclareMathOperator*{\argmin}{arg\,min}

\begin{document}

\title{Threshold-Based Fast Successive-Cancellation Decoding of Polar Codes}

\author{Haotian~Zheng,~\IEEEmembership{Student Member,~IEEE}, Seyyed~Ali~Hashemi,~\IEEEmembership{Member,~IEEE}, Alexios~Balatsoukas-Stimming,~\IEEEmembership{Member,~IEEE}, Zizheng~Cao,~\IEEEmembership{Member,~IEEE}, Ton~Koonen,~\IEEEmembership{Fellow,~IEEE}, John~Cioffi,~\IEEEmembership{Fellow,~IEEE}, Andrea~Goldsmith,~\IEEEmembership{Fellow,~IEEE}
\thanks{H.~Zheng, A.~Balatsoukas-Stimming, Z.~Cao, and A.~M.~J.~Koonen are with the Department of Electrical Engineering, Eindhoven University of Technology, The Netherlands (\mbox{e-mails:} \{h.zheng, a.k.balatsoukas.stimming, z.cao, a.m.j.koonen\}@tue.nl). 
}
\thanks{S.~A.~Hashemi, J.~Cioffi, and A.~Goldsmith are with the Department of Electrical Engineering, Stanford University, Stanford, CA 94305 USA (\mbox{e-mails:} \{ahashemi, cioffi\}@stanford.edu, andrea@wsl.stanford.edu). (Corresponding author: Zizheng Cao)}

\thanks{Part of this work was presented at the IEEE International Conference on Communications, June 2020.}
}

\maketitle

\vspace{-3em}
\begin{abstract}
Fast SC decoding overcomes the latency caused by the serial nature of the SC decoding by identifying new nodes in the upper levels of the SC decoding tree and implementing their fast parallel decoders. In this work, we first present a novel sequence repetition node corresponding to a particular class of bit sequences. Most existing special node types are special cases of the proposed sequence repetition node. Then, a fast parallel decoder is proposed for this class of node. To further speed up the decoding process of general nodes outside this class, a threshold-based hard-decision-aided scheme is introduced. The threshold value that guarantees a given error-correction performance in the proposed scheme is derived theoretically. Analysis and hardware implementation results on a polar code of length $1024$ with code rates $1/4$, $1/2$, and $3/4$ show that our proposed algorithm reduces the required clock cycles by up to $8\%$, and leads to a $10\%$ improvement in the maximum operating frequency compared to state-of-the-art decoders without tangibly altering the error-correction performance. In addition, using the proposed threshold-based hard-decision-aided scheme, the decoding latency can be further reduced by $57\%$ at $\mathrm{E_b}/\mathrm{N_0} = 5.0$~dB.
\end{abstract}

 \begin{IEEEkeywords}
 Polar codes, Fast successive-cancellation decoding, Sequence repetition node, 
 Threshold-based hard-decision-aided scheme.
 \end{IEEEkeywords}

\IEEEpeerreviewmaketitle

\section{Introduction}
\label{sec:intro}

\IEEEPARstart{P}{olar} codes represent a channel coding scheme that can provably achieve the capacity of binary-input memoryless channels \cite{arikan2009}. The explicit coding structure and low-complexity successive-cancellation (SC) decoding algorithm has generated significant interest in polar code research across both industry and academia. In particular, in the latest cellular standard of 5G \cite{3GPP}, polar codes are adopted in the control channel of the enhanced mobile broadband (eMBB) use case. Although SC decoding provides a low-complexity capacity-achieving solution for polar codes with long block length, its sequential bit-by-bit decoding nature leads to high decoding latency, which constrains its application in low-latency communication scenarios such as the ultra-reliable low-latency communication (URLLC) \cite{3GPP} scheme of 5G. Therefore, the design of fast SC-based decoding algorithms for polar codes with low decoding latency has received a lot of attention~\cite{niu2014polar}.

A look-ahead technique was adopted to speed up SC decoding in \cite{zhang2012reduced,BYuan2Bit,ASIC} by pre-computing all the possible likelihoods of the bits that have not been decoded yet, and selecting the appropriate likelihood once the corresponding bit is estimated. Using the binary tree representation of SC decoding of polar codes, instead of working at the bit-level that corresponds to the leaf nodes of the SC decoding tree, parallel multi-bit decision is performed at the intermediate nodes of the SC decoding tree. An exhaustive-search decoding algorithm is used in \cite{yuan2015reduced,yuan2015low,sarkis2013increasing,husmann2017reduced} to make multi-bit decisions and to avoid the latency caused by the traversal of the SC decoding tree to compute the intermediate likelihoods. However, due to the high complexity of the exhaustive search, this method is generally only suitable for nodes that represent codes of very short lengths.

It was shown in \cite{alamdar2011simplified} that a node in the SC decoding tree that represents a code of rate $0$ (Rate-0 node) or a code of rate $1$ (Rate-1 node) can be decoded efficiently without traversing the SC decoding tree. In \cite{sarkis2014fast}, fast decoders of repetition (REP) and single parity-check (SPC) nodes were proposed for the SC decoding. Using these nodes, a fully-unrolled hardware architecture was proposed in \cite{giard2016multi}. Techniques were developed in \cite{huang2012latency,balatsoukas2014enabling,zhang2015simplified,638mbps,giard2018fast} to adjust the codes that are represented by the nodes in the SC decoding tree to increase the number of nodes that can be decoded efficiently. However, these methods result in a degraded error-correction performance. On the basis of the works in \cite{alamdar2011simplified,sarkis2014fast}, five new nodes (Type-I, Type-II, Type-III, Type-IV, and Type-V) were identified and their fast decoders were designed in \cite{hanif2017fast}. In \cite{condo2018generalized}, a generalized REP (G-REP) node and a generalized parity-check (G-PC) node were proposed to reduce the latency of the SC decoding even further. In \cite{gamage2019low}, seven of the most prevalent node patterns in short polar codes were analysed and efficient algorithms for processing these node patterns in parallel are proposed. However, the decoding of some of these node patterns leads to significant performance loss. 
Memory footprint optimization and operation merging were described in \cite{ercan2017reduced,ercan2019operation} to improve the speed of SC decoding. The combination of fast SC and fast SC flip decoding algorithms was investigated in \cite{GiardWCNCW,Ercan2020}. Moreover, the identification and the utilization of the aforementioned special nodes were also extended to SC list (SCL) \cite{tal2015list} decoding \cite{sarkis2016fast,hashemi2016fast,Hashemi2017,Hashemi2019,HanifTOC}. All these works require the design of a separate decoder for each class of node, which inevitably increases the implementation complexity. In addition, as shown in this work, the achievable parallelism in decoding can be further increased without degrading the error-correction performance.

For general nodes that do not fall in one of the above node categories, \cite{li2018low} proposed a hard-decision scheme based on node error probability. Specifically, in that work it was shown that extra latency reduction can be achieved when the communications channel has low noise. However, the hard-decision threshold is calculated empirically rather than for a desired error-correction performance. In \cite{sun2019}, a hypothesis-testing-based
strategy is designed to select reliable unstructured nodes for hard decision. However, additional operations are required to be performed to calculate the decision rule, thus, incurring extra decoding latency. For all the existing hard-decision schemes, a threshold comparison operation is required each time a general constituent code is encountered in the course of the SC decoding algorithm.

In this paper, a fast SC decoding algorithm with a higher degree of parallelism than the state of the art is proposed.
First, a class of \emph{sequence repetition} (SR) nodes is proposed which provides a unified description of most of the existing special nodes. This class of nodes is typically found at a higher level and has a higher frequency of occurrence in the decoding tree than other existing special nodes. 5G polar codes with different code lengths and code rates can be represented using only SR nodes. Utilizing this class of nodes, a simple and efficient fast simplified SC decoding algorithm called the \emph{SR node-based fast SC} (SRFSC) decoding algorithm is proposed, which achieves a high degree of parallelism without degrading the error-correction performance. Employing only SR nodes, the SRFSC decoder requires a lower number of time steps than most existing works. Moreover, SR nodes can be implemented together with other operation mergers, such as node-branch and branch mergers, to outperform the state-of-the-art decoders in terms of the required number of time steps. In addition, if a realistic hardware implementation is considered, the proposed SRFSC decoder provides up to $8\%$ reduction in terms of the required number of clock cycles and $10\%$ improvement in terms of the maximum operating frequency with respect to the state-of-the-art decoders on a polar codes of length $1024$ changed{with code rates $1/4$, $1/2$, and $3/4$.

Second, a threshold-based hard-decision-aided (TA) scheme is proposed to speed up the decoding of the nodes that are not SR nodes for a binary additive white Gaussian noise (BAWGN) channel. Consequently, a TA-SRFSC decoding algorithm is proposed that adopts a simpler threshold for hard-decision than that in \cite{li2018low}. The effect of the defined threshold on the error-correction performance of the proposed TA-SRFSC decoding algorithm is analyzed. Moreover, a systematic way to derive the threshold value for a desired upper bound for its block error rate (BLER) is determined. Performance results show that, with the help of the proposed TA scheme, the decoding latency of SRFSC decoding can be further reduced by $57\%$ at $\mathrm{E_b}/\mathrm{N_0}=5$~dB on a polar code of length $1024$ and rate $1/2$. In addition, a multi-stage decoding strategy is introduced to mitigate the possible error-correction performance loss of TA-SRFSC decoding with respect to SC decoding, while achieving average decoding latency comparable to TA-SRFSC decoding.

The rest of this paper is organized as follows. Section~\ref{sec:pre} gives a brief introduction to the basic concept of polar codes and fast SC decoding. In Section~\ref{sec:sr}, the SRFSC decoding algorithm is introduced. With the help of the proposed TA scheme, the TA-SRFSC decoding is presented in Section~\ref{sec:TA}. Section~\ref{sec:performance} analyzes the decoding latency and simulation results are shown in Section~\ref{sec:Numerical results}. Finally, Section~\ref{sec:Conclu} gives a summary of the paper and concluding remarks.

\section{Preliminaries}
\label{sec:pre}

\subsection{Notation Conventions}
In this paper, blackboard letters, such as $\mathbb{X}$, denote a set and $\left|\mathbb{X}\right|$ denotes the number of elements in $\mathbb{X}$. Bold letters, such as $\boldsymbol v$, denote a row vector, $\boldsymbol{v}^T$ denotes the transpose of $\boldsymbol v$, and notation $v\left[i:j\right]$, $1\leq i<j$ represents a subvector $\left(v\left[i\right],v\left[i+1\right],\dots,v\left[j\right]\right)$. $\oplus$ is used as the bitwise XOR operation and $v\left[i:j\right]\oplus z=\left(v\left[i\right]\oplus z,v\left[i+1\right]\oplus z,\dots,v\left[j\right]\oplus z\right)$, $z\in\left\{0,1\right\}$. The Kronecker product of two matrices $\mathbf{F}$ and $\mathbf{G}$ is written as $\displaystyle \mathbf F\otimes \mathbf G$.
${\mathbf F}_N$ represents an $N\times N$ square matrix and $\mathbf F_N^{\otimes n}$ denotes the $n$-th Kronecker power of $\mathbf F_N$. Throughout this paper, $\ln(x)$ denotes the natural logarithm and $\log(x)$ indicates the base-$2$ logarithm of $x$, respectively.  

\vspace{-1em}
\subsection{Polar Codes}
\label{sec:PC}
A polar code with code length $N=2^n$ and information length $K$ is denoted by $\mathcal{P}\left(N,K\right)$ and has rate $R = K/N$. The encoding can be expressed as $\boldsymbol{x}=\boldsymbol{u}\mathbf{G}_N$, where $\boldsymbol{u}=\left(u\left[1\right],u\left[2\right],\ldots,u\left[{N}\right]\right)$ is the input bit sequence and $\boldsymbol{x}=\left(x\left[1\right],x\left[2\right],\ldots,x\left[{N}\right]\right)$ is the encoded bit sequence. $\mathbf{G}_N=\mathbf{R}_N\mathbf{F}_2^{\otimes n}$ is the generator matrix, where $\mathbf{R}_N$ is a bit-reversal permutation matrix and $\mathbf{F}_2=\left[\begin{smallmatrix}1&0\\1&1\end{smallmatrix}\right]$.

The input bit sequence $\boldsymbol{u}$ consists of $K$ information bits and $N-K$ frozen bits. The information bits form set $\mathbb{A}$, transmitting information bits, while the frozen bits form set $\mathbb{A}^c$, transmitting fixed bits known to the receiver. For symmetric channels, without loss of generality, all frozen bits are set to zero \cite{arikan2009}. To distinguish between frozen and information bits, a vector of flags $\boldsymbol{d} = \left(d\left[1\right],d\left[2\right],\ldots,d\left[{N}\right]\right)$ is used where each flag $d\left[k\right]$ is assigned as
\begin{equation} \label{eq:flag}
d\left[k\right] = \begin{cases} 0, &\mbox{if } k\in \mathbb{A}^c , \\ 
1, & \mbox{otherwise.} \end{cases}
\end{equation}
The codeword $\boldsymbol{x}$ is transmitted through a channel after modulation. In this paper, non-systematic polar codes and binary phase-shift keying (BPSK) modulation that maps $\left\{0,1\right\}$ to $\left\{+1,-1\right\}$ are considered. Transmission takes place over an additive white Gaussian noise (AWGN) channel.

\vspace{-1em}
\subsection{SC Decoding and Binary Tree Representation}
\label{sec:tree}

\begin{figure}[htbp]
\begin{minipage}[t]{0.5\textwidth}
\centering
\scalebox{0.6}{\includegraphics{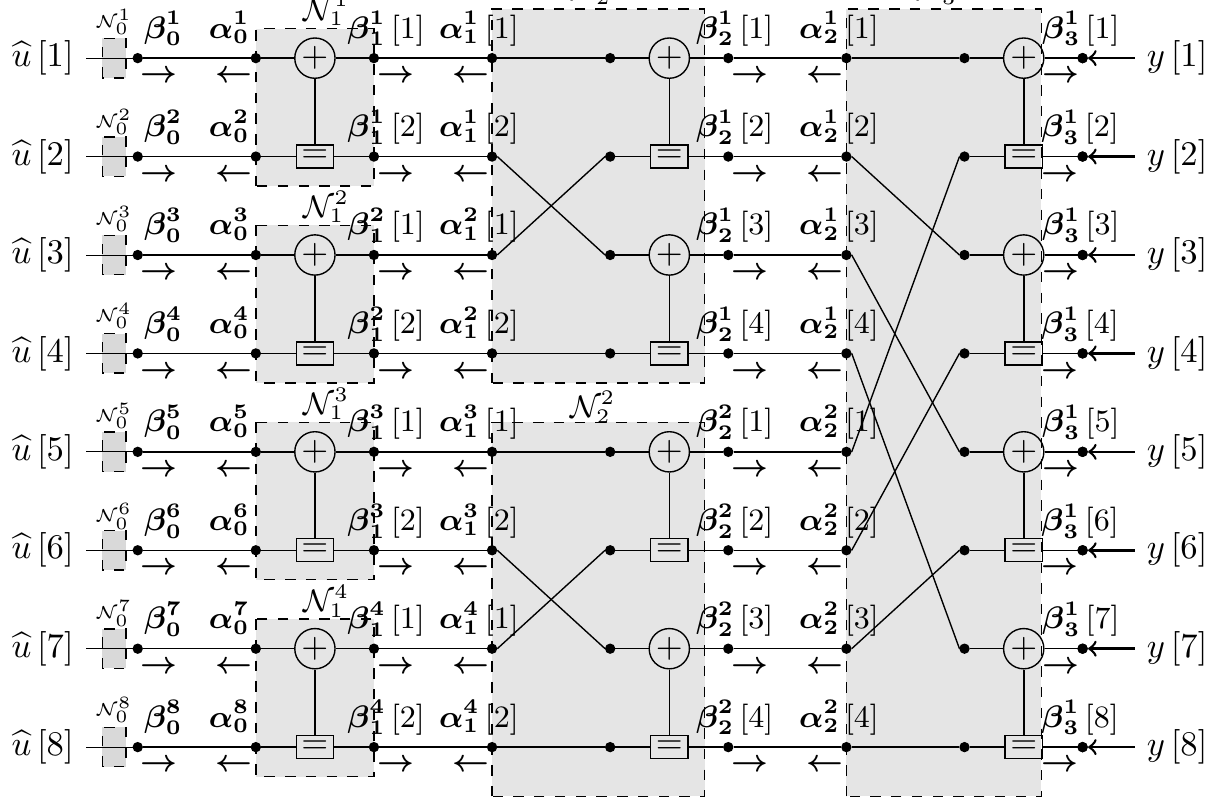}}
\caption{SC decoding on the factor graph of a polar code $N=8$.}
\label{fig:graph}
\end{minipage}

\vspace{1em}
\begin{minipage}[t]{0.5\textwidth}
\centering
\scalebox{0.6}{\includegraphics{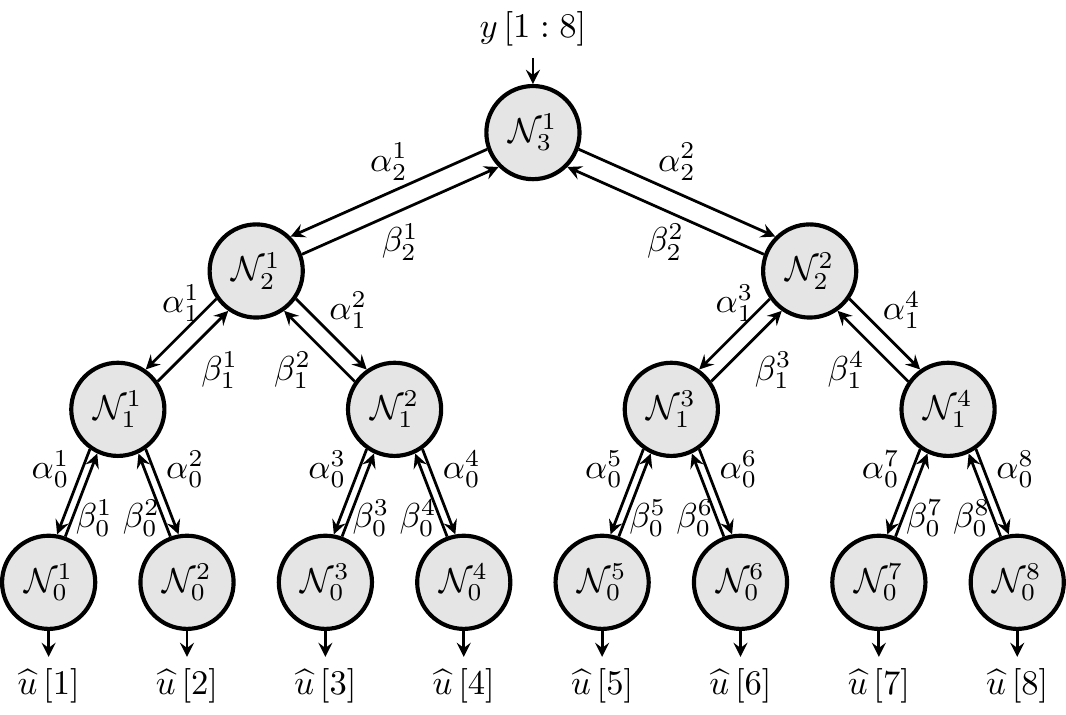}}
\caption{Binary tree representation of a SC decoder for a polar code with $N=8$.}
\label{fig:tree}
\end{minipage}
\end{figure}

SC decoding can be illustrated on the factor graph of polar codes as shown in Fig.~\ref{fig:graph}. The factor graph consists of $n+1$ levels and, by grouping all the operations that can be performed in parallel, SC decoding can be represented as the traversal of a binary tree. This traversal is shown in Fig.~\ref{fig:tree}, starting from the left side of the binary tree. At level $j$ of the SC decoding tree with $n+1$ levels, there are $2^{n-j}$ nodes ($0\leq j \leq n$), and the $i$-th node at level $j$ ($1\leq i\leq 2^{n-j}$) of the SC decoding tree is denoted as $\mathcal{N}_j^i$. The left and the right child nodes of $\mathcal{N}_j^i$ are $\mathcal{N}_{j-1}^{2i-1}$ and $\mathcal{N}_{j-1}^{2i}$, respectively, as illustrated in Fig.~\ref{fig:tree}. For $\mathcal{N}_j^i$, $\alpha_j^i\left[k\right]$, $1\leq k\leq2^j$, indicates the $k$-th input logarithmic likelihood ratio (LLR) value, and $\beta_j^i\left[k\right]$, $1\leq k\leq2^j$, denotes the $k$-th output binary hard-valued message. For the AWGN channel, the received vector $\boldsymbol{y} = \left(y\left[1\right],y\left[2\right],\ldots,y\left[N\right]\right)$ from the channel can be used to calculate the channel LLR as $2\boldsymbol y/\sigma^2$, where $\sigma^2$ is the variance of the Gaussian noise. SC decoding starts by setting $\alpha_n^1\left[1:N\right] = 2\boldsymbol y/\sigma^2$. A node will be activated once all its inputs are available. When LLR messages pass through a node in the factor graph which is indicated by the $\oplus$ sign, the $f$ function over the LLR domain is executed as
\begin{equation} \label{eq:f_function}
\alpha_{j-1}^{2i-1}\left[k\right]=f\left(\alpha_j^{i}\left[{2k-1}\right],\alpha_j^{i}\left[{2k}\right]\right),
\end{equation}
and when LLR messages pass through a node in the factor graph which is indicated by the {\scriptsize$\raisebox{1.6pt}{\boxed{=}}$} sign, the $g$ function over the LLR domain is executed as
\begin{equation}
\label{eq:g_function}
\alpha_{j-1}^{2i}\left[k\right]=g\left(\alpha_j^{i}\left[{2k-1}\right],\alpha_j^{i}\left[{2k}\right],\beta_{j-1}^{2i-1}\left[k\right]\right),
\end{equation}
where
\begin{align}
\label{eq:F} f\left(x,y\right) &= 2\arctanh\left(\tanh\left(\frac x2\right)\tanh\left(\frac y2\right)\right),\\
\label{eq:G} g\left(x,y,u\right) &= \left(-1\right)^ux+y.
\end{align}
The $f$ function can be approximated as \cite{leroux2011hardware}
\begin{equation}
f\left(x,y\right) = \mathrm{sign}\left(x\right)\mathrm{sign}\left(y\right)\min\left(\left|x\right|,\left|y\right|\right).
\end{equation}
When the LLR value of the $k$-th bit at level zero $\alpha_0^k,\;1\leq k\leq N$, is calculated, the estimation of $u\left[k\right]$, denoted as ${\hat u}\left[k\right]$, can be obtained as
\begin{equation}\label{eq:decision}
\centering      
\hat{u}\left[k\right]=\hat\beta_0^k=\begin{cases} 0, &\mbox{if } k\in \mathbb{A}^c \text{,} \\ 
\frac{1-\mathrm{sign}(\alpha_0^k)}{2}, & \mbox{otherwise.} \end{cases}
\end{equation}
The hard-valued messages are propagated back to the parent node as
\begin{equation}
\hat\beta_j^i\left[k\right]=
\begin{cases} \hat\beta_{j-1}^{2i-1}\left[\frac{k+1}2\right]\oplus\hat\beta_{j-1}^{2i}\left[\frac{k+1}2\right], &\mbox{if} \mod(k,2)=1 \text{,} \\ 
\hat\beta_{j-1}^{2i}\left[\frac{k}2\right], & \mbox{if} \mod(k,2)=0. \end{cases}
\label{eq:hardpropagation}
\end{equation}
After traversing all the nodes in the SC decoding tree, $\hat{\boldsymbol{u}}$ contains the decoding result. Thus the latency of SC decoding for a polar code of length $N$ in terms of the number of time steps can be represented by the number of nodes in the SC decoding tree as \cite{arikan2009}
\begin{equation}
    {\mathcal{T}}_{\text{SC}} = 2N-2 \text{.}
\end{equation}

\vspace{-1em}
\subsection{Fast SC Decoding}
\label{sec:FSC}

The SC decoding has strong data dependencies that limit the amount of parallelism that can be exploited within the algorithm because the estimation of each bit depends on the estimation of all previous bits. This leads to a large latency in the SC decoding algorithm.
It was shown in \cite{sarkis2013increasing} that for a node $\mathcal{N}_j^i$, $\beta_j^i\left[1:2^j\right]$ can be estimated without traversing the decoding tree as
\begin{equation} \label{eq:estimate}
\hat\beta_j^i\left[1:2^j\right]= \underset{\beta_j^i\left[1:2^j\right]\in\mathbb{C}_j^i}{\argmax}\sum_{k=1}^{2^j}\left(-1\right)^{\beta_j^i\left[k\right]}\alpha_j^i\left[k\right],
\end{equation}
where $\mathbb{C}_j^i$ is the set of all the codewords associated with node $\mathcal{N}_j^i$.
Multibit decoding can be performed directly in an intermediate level instead of bit-by-bit sequential decoding at level $0$, in order to traverse fewer nodes in the SC decoding tree and consequently, to reduce the latency caused by data computation and exchange. However, the evaluation of (\ref{eq:estimate}) generally requires exhaustive search over all the codewords in the set $\mathbb{C}_j^i$ which is computationally intensive in practice. In \cite{sarkis2014fast}, a fast SC (FSC) decoding algorithm was proposed which performs fast parallel decoding when specific special node types are encountered. The sequences of information and frozen bits of these node types have special bit-patterns. Therefore, they can be decoded more efficiently without the need for exhaustive search. Using the vector $\boldsymbol d$, the five special node types proposed in \cite{sarkis2014fast} are described as:
\begin{itemize}
\item Rate-0 node: all bits are frozen bits, $\boldsymbol d=\left(0,0,...,0\right)$.
\item Rate-1 node: all bits are non-frozen bits, $\boldsymbol d=\left(1,1,...,1\right)$.
\item REP node: all bits are frozen bits except the last one, $\boldsymbol d=\left(0,...,0,1\right)$.
\item SPC node: all bits are non-frozen bits except the first one, $\boldsymbol d=\left(0,1,...,1\right)$.
\item ML node: a code of length $4$ with $\boldsymbol d=\left(0,1,0,1\right)$.
\end{itemize}
By merging the aforementioned special nodes, five additional nodes were proposed, namely, the REP-SPC node, which is a REP node followed by a SPC node; the P-01/P-0SPC node, which is generated by merging a Rate-0 node with a Rate-1/SPC node; and the P-R1/P-RSPC node, which is generated by merging a $g$ function branch operation in~\eqref{eq:G} with the following Rate-1/SPC node.
In \cite{hanif2017fast}, five additional special node types and their corresponding fast decoders were introduced. This enhanced FSC decoding algorithm can achieve a lower decoding latency than FSC decoding. The five special node types are:
\begin{itemize}
\item Type-I node: all bits are frozen bits except the last two, $\boldsymbol d=\left(0,...,0,1,1\right)$.\par
\item Type-II node: all bits are frozen bits except the last three, $\boldsymbol d=\left(0,...,0,1,1,1\right)$.\par
\item Type-III node: all bits are non-frozen bits except the first two, $\boldsymbol d=\left(0,0,1,...,1\right)$.\par
\item Type-IV node: all bits are non-frozen bits except the first three, $\boldsymbol d=\left(0,0,0,1,...,1\right)$.\par
\item Type-V node: all bits are frozen bits except the last three and the fifth to last, $\boldsymbol d=\left(0,...,0,1,0,1,1,1\right)$.\par
\end{itemize}

A generalized FSC (GFSC) decoding algorithm was proposed in \cite{condo2018generalized} by introducing the G-PC node and the G-REP node (previously identified as Multiple-G0 nodes in \cite{ercan2019operation}). The G-PC node is a node at level $j$ having all its descendants as Rate-1 nodes except the leftmost one at a certain level $r<j$, that is a Rate-0 node. The G-REP node is a node at level $j$ for which all its descendants are Rate-0 nodes, except the rightmost one at a certain level $r<j$, which is a generic node of rate $C$ (Rate-C). 

In \cite{giard2018fast}, in addition to the nodes in \cite{sarkis2014fast}, three special node types are added: the REP1 node, with the REP node on the left and the Rate-1 node on the right; the 0REPSPC node, which is a concatenation of the Rate-0, REP, and SPC nodes; and the 001 node, whose left $3/4$ of bits are frozen bits and right $1/4$ of bits are unfrozen bits.

To further reduce the latency, the operation merging scenarios are generalized in \cite{ercan2017reduced} and \cite{ercan2019operation} where, in addition to merging special nodes, branch operation (\eqref{eq:F},\eqref{eq:G} and \eqref{eq:hardpropagation}) merging is also considered. Consequently, REP-REPSPC, REP-Rate1, and Rate0-ML nodes are adopted as the merging of special nodes, and F-REP node is adopted as a node-branch merger, which is generated by performing an $f$ function operation followed by a REP node. The following branch operation mergers are further introduced: 

 \begin{itemize}
 \item $\text{F}^{\times2}$: two consecutive $f$ function operations in~\eqref{eq:F}.
 \item $\text{G0}^{\times2}$: two consecutive $\text{G0}$ operations, where $\text{G0}$ is the $g$ function in~\eqref{eq:G} assuming $u=0$.
 \item C$^{\times2}$, C$^{\times3}$: up to three consecutive operations in~\eqref{eq:hardpropagation}.
 \item C0$^{\times2}$, C0$^{\times3}$: up to three consecutive operations in~\eqref{eq:hardpropagation}, assuming the estimations from the left branch are all zeros.
 \item $\text{G-F}$: $g$ function operation followed by an $f$ function operation.
 \item $\text{F-G0}$: $f$ function operation followed by a $\text{G0}$ operation.\par
 \end{itemize}

The key advantage of using specific parallel decoders for the aforementioned special nodes is that, since the SC decoding tree is not traversed when one of these nodes is encountered, significant latency saving can be achieved. For example, if $\mathcal{N}_j^i$ is a Rate-1 node, hard decision decoding can be used to immediately obtain the decoding result as
\begin{equation}
\hat\beta_j^i\left[k\right]=h\left(\alpha_j^i\left[k\right]\right)=
\begin{cases} 0, &\mbox{if } \alpha_j^i\left[k\right]\geq0 \text{,} \\ 
1, & \mbox{otherwise.} \end{cases}
\label{eq:rate1}
\end{equation}
If $\mathcal{N}_j^i$ is a SPC node, hard decision based on (\ref{eq:rate1}) is first derived followed by the calculation of the parity of the output using modulo-2 addition. The index of the least reliable bit is found as
\begin{equation}
k'=\underset k{\argmin}\left|\alpha_j^i\left[k\right]\right|.
\end{equation}
Finally, the bits in a SPC node are estimated as
\begin{equation}
\hat\beta_j^i\left[k\right]=
\begin{cases} h\left(\alpha_j^i\left[k\right]\right)\oplus\mathrm{parity}, &\mbox{if } k = k' \text{,} \\ 
h\left(\alpha_j^i\left[k\right]\right), & \mbox{otherwise.} \end{cases}
\end{equation}
This operation can be performed in a single time step \cite{hanif2017fast}. Finally, if $\mathcal{N}_j^i$ is a G-PC node, the decoding can be viewed as a parallel decoding of several separate SPC nodes. The decoding of a G-PC node can generally be performed in one time step considering parallel SPC decoders~\cite{condo2018generalized}.

All of the aforementioned fast SC decoding algorithms perform  decoding at an intermediate level of the decoding tree in order to reduce the number of traversed nodes. A decoding algorithm that can decode a node at a higher level of the decoding tree generally results in more savings in terms of latency than one that decodes a node at a lower level of the decoding tree.

\vspace{-1em}
\subsection{Threshold-based Hard-decision-aided Scheme}
\label{sec:OLDTA}
For a  binary AWGN channel with standard deviation $\sigma_n$, it was shown in \cite{trifonov2012efficient} that, considering all the previous bits are decoded correctly, the LLR value $a_j^i\left[k\right]$, $1\le k\le 2^j$, input into node $\mathcal{N}_j^i$ can be approximated as a Gaussian variable using a Gaussian approximation as
\begin{equation} \label{eq:gaussian}
\alpha_j^i\left[k\right]\sim \mathcal{N}\left(M_j^i\left[k\right],2\left|M_j^i\left[k\right]\right|\right),
\end{equation}
where $M_j^i\left[k\right]$ is the expectation of $\alpha_j^i\left[k\right]$ \cite{zhang2016split} such that
\begin{equation} \label{eq:M}
M_j^i\left[k\right]=
\begin{cases} m_j^i, & \text{if } \beta_j^i\left[k\right]=0 \text{,} \\ 
-m_j^i, & \text{if } \beta_j^i\left[k\right]=1, \end{cases}
\end{equation}
and $m_j^i$ can be calculated recursively offline assuming the all-zero codeword is transmitted as
\begin{equation} \label{eq:recursive1}
m_n^{1}=2/\sigma_n^2,\;\;m_{j-1}^{2i-1}=\varphi^{-1}(1-\lbrack1-\varphi(m_j^{i})\rbrack^2),\;\;m_{j-1}^{2i}=2m_j^{i},
\end{equation}
where
\begin{equation} \label{eq:varphi}
\varphi(x)=
\begin{cases} 1-\frac1{\sqrt{4|x|}}\int_{-\infty}^\infty \tanh\left(\frac{u}{2}\right) e^{-\frac{(u-x)^2}{4|x|}}du, & x\neq 0 \text{,} \\
0, & x=0. \end{cases}
\end{equation}

It was shown in \cite{li2018low} that, when the magnitude of the LLR values at a certain node in the SC decoding tree is large enough, the node has enough reliability to perform hard decision directly at the node without tangibly altering the error-correction performance. To determine the reliability of the node, a threshold is defined in \cite{li2018low} as
\begin{equation}
\label{eq:oldthreshold} 
T=c_t\log\frac{1-\frac12\mathrm{erfc}\left(0.5\sqrt{m_j^i}\right)}{\frac12\mathrm{erfc}\left(0.5\sqrt{m_j^i}\right)},
\end{equation}
where $c_t\geq1$ is a constant that is selected empirically. The hard-decision estimate of the received LLR values is calculated using
\begin{equation} \label{eq:hard decision}
\mathrm{HB}_j^i\left[k\right]=
\begin{cases} 0, & \text{if }\alpha_j^i\left[k\right]>T, \\ 
1, & \text{if }\alpha_j^i\left[k\right]<-T. \end{cases}
\end{equation}

The issue with the method in \cite{li2018low} is that the threshold defined in (\ref{eq:oldthreshold}) contains complex calculations of complementary error functions $\mathrm{erfc}\left(\cdot\right)$, making the corresponding calculation inefficient. Moreover, the hard-decision threshold is calculated empirically rather than for a desired error-correction performance and the threshold comparison in (\ref{eq:hard decision}) is performed every time a node with no specific structure is encountered in the SC decoding process.

\section{Fast SC Decoding with Sequence Repetition Nodes}
\label{sec:sr}

This section introduces a class of nodes that is at a higher level of the SC decoding tree and shows how parallel decoding at this class of nodes can be exploited to achieve significant latency savings in comparison with the state of the art.
\subsection{Sequence Repetition (SR) Node}
\label{sec:sr node}

Let $\mathcal{N}_j^i$ be a node at level $j$ of the tree representation of SC decoding as shown in Fig.~\ref{fig:tree}. An SR node is any node at stage $j$ whose descendants are all either Rate-0 or REP nodes, except the rightmost one at a certain stage $r$, $0\leq r\leq j$, that is a generic node of rate $C$. The structure of an SR node is depicted in Fig.~\ref{fig:SR}.
The rightmost node $\mathcal{N}_r^{i\times 2^{j-r}}$ at stage $r$ is denoted as the source node of the SR node $\mathcal{N}_j^i$. Let $E = i\times 2^{j-r}$ so the source node can be denoted as $\mathcal{N}_r^E$.

An SR node can be represented by three parameters as $\text{SR}(\boldsymbol{v},\text{SNT},r)$, where $r$ is the level of the SC decoding tree in which $\mathcal{N}_r^E$ is located, SNT is the source node type, and $\boldsymbol{v} = \left(v\left[{j}\right],v\left[{j-1}\right],\ldots,v\left[{r+1}\right]\right)$ is a vector of length $\left(j-r\right)$ such that for the left child node of the parent node of $\mathcal{N}_r^E$ at level $k$, $r<k\leq j$, $v\left[{k}\right]$ is calculated as
\begin{equation}
v\left[k\right] = \begin{cases}
0, & \text{if the left child node is a Rate-0 node,}\\
1, & \text{if the left child node is a REP node.}
\end{cases}
\end{equation}
Note that when $r=j$, $\mathcal{N}_j^i$ is a source node and thus $\boldsymbol{v}$ is an empty vector denoted as $\boldsymbol{v} = \emptyset$.


\begin{figure}[htbp]
\begin{minipage}[t]{0.49\textwidth}
\centering
\scalebox{0.85}{\includegraphics{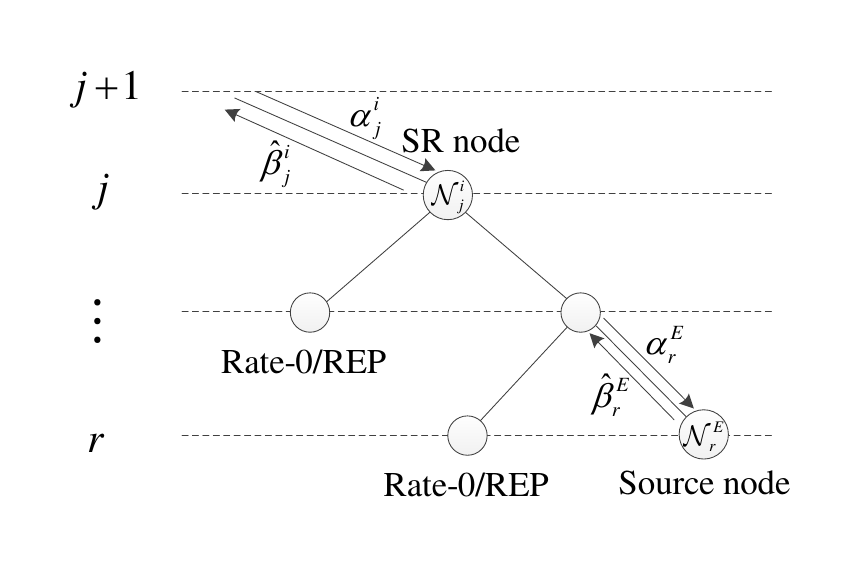}}
\setlength{\abovecaptionskip}{-18pt}
\caption{General structure of SR Node.}
\label{fig:SR}
\end{minipage}
\hfill
\begin{minipage}[t]{0.5\textwidth}
\centering
\scalebox{0.85}{\includegraphics{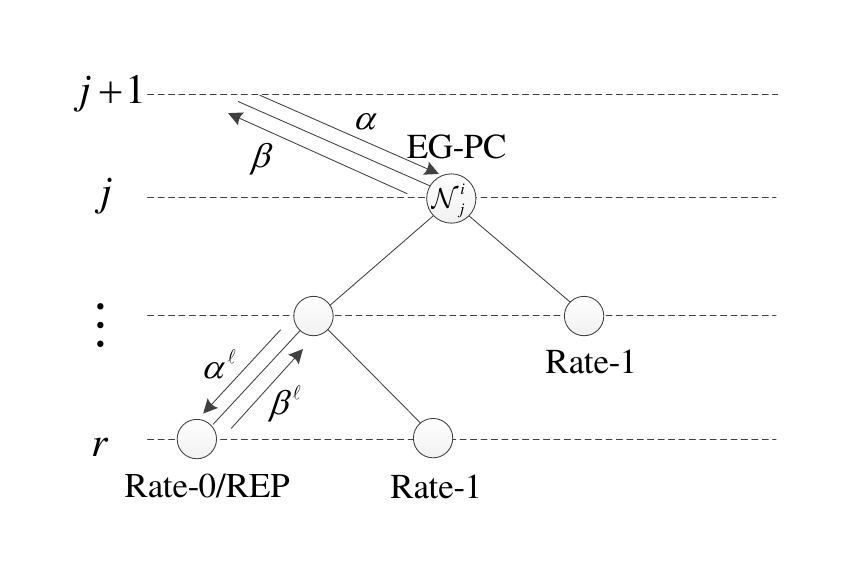}}
\setlength{\abovecaptionskip}{-18pt}
\caption{General structure of Extended G-PC Node.}
\label{fig:EG-PC}
\end{minipage}
\end{figure}

\subsection{Source Node}
\label{sec:source node}

To define the source node type, an extended class of G-PC (EG-PC) nodes is first introduced. The structure of the EG-PC node is depicted in Fig.~\ref{fig:EG-PC}. The EG-PC node is different from the G-PC node in its leftmost descendant node that can be either a Rate-0 or a REP node. The bits in an EG-PC node satisfy the following parity check constraint,
\begin{equation}
z = \bigoplus_{m=\left(k-1\right)2^{j-r}+1}^{k\times2^{j-r}}\beta_j^i\left[m\right],
\end{equation}
where $k\in\left\{1,\ldots,2^r\right\}$, and $z\in\left\{0,1\right\}$ is the parity. Unlike G-PC nodes whose parity is always even ($z=0$), the EG-PC node can have either even parity ($z=0$) or odd parity ($z=1$). The parity of the EG-PC node can be calculated as
\begin{equation}
z = \begin{cases}
0, & \text{if the leftmost node is Rate-0,}\\
h\left(\sum_{k=1}^{2^r}\alpha_k\right), & \text{otherwise,}
\end{cases}
\end{equation}
where $\alpha_k=2\tanh^{-1}\left(\prod_{m=\left(k-1\right)2^{j-r}+1}^{k2^{j-r}}\tanh\left(\frac{\alpha_j^i\left[m\right]}2\right)\right)$.
After computing $z$, Wagner decoders \cite{silverman1954coding}
can be used to decode the $2^r$ SPC codes with either even or odd parity constraints. A Wagner decoder performs hard decisions on the bits, and if the parity does not hold, it flips the bit with the lowest absolute LLR value.

SPC, Type-III, Type-IV, and G-PC nodes can be represented as special cases of EG-PC nodes. As a result, most of the common special nodes can be represented as SR nodes, as shown in Table~\ref{tab:SR}.
Note that the node-branch mergers like P-RSPC and F-REP, and branch operation mergers such as $\text{F}^{\times2}$, $\text{G‐F}$, $\text{G0}^{\times2}$, $\text{F‐{G0}}$, do not fit into the category of SR nodes. It is worth mentioning that other parameter choices for SR nodes may lead to the discovery of more types of special nodes that can be decoded efficiently, but this exploration is beyond the scope of this work.

\begin{table}
\def\arraystretch{0.7}
\centering
\caption{SR node representation of common node types.}
\setlength{\belowcaptionskip}{-10pt}
\vspace{-1em}
\begin{tabular}{llllll}
\toprule
Node Type & SR Node Representation & Length of $\boldsymbol v$ \\
\midrule
Rate-0		& $\mathrm{SR}(\emptyset,\text{Rate-0},j)$ & 0 \\
 REP			& $\mathrm{SR}((0,\ldots,0),\text{Rate-1},0)$ & $j$\\
SPC			& $\mathrm{SR}(\emptyset,\text{EG-PC},j)$ & 0 \\
 Rate-1		& $\mathrm{SR}(\emptyset,\text{Rate-1},j)$ & 0\\
P-01		& $\mathrm{SR}((0),\text{Rate-1},j-1)$ & $1$ \\
 P-0SPC	& $\mathrm{SR}((0),\text{EG-PC},j-1)$ & $1$\\
Type-I		& $\mathrm{SR}((0,\ldots,0),\text{Rate-1},1)$ & $j-1$ \\
 Type-II		& $\mathrm{SR}((0,\ldots,0),\text{EG-PC},2)$ & $j-2$ \\
Type-III	& $\mathrm{SR}(\emptyset,\text{EG-PC},j)$ & 0 \\
 Type-IV		& $\mathrm{SR}(\emptyset,\text{EG-PC},j)$ & 0\\
Type-V		& $\mathrm{SR}((0,\ldots,0,1),\text{EG-PC},2)$ & $j-2$ \\
 G-PC		& $\mathrm{SR}(\emptyset,\text{EG-PC},j)$ & 0\\
G-REP		& $\mathrm{SR}((0,\ldots,0),\text{Rate-C},r)$ & $j-r$ \\
 REP-SPC	& $\mathrm{SR}((1),\text{EG-PC},2)$ & $1$\\
0REPSPC		& $\mathrm{SR}((0,1),\text{EG-PC},j-2)$ & $2$ \\
 001	& $\mathrm{SR}((0,0),\text{Rate-1},j-2)$ & $2$\\
REP-REPSPC	& $\mathrm{SR}((1,1),\text{EG-PC},j-2)$ & $2$ \\
 Rate0-ML	& $\mathrm{SR}((0,1,0),\text{Rate-1},0)$ & $3$\\
REP-Rate1(REP1)	& $\mathrm{SR}((1),\text{Rate-1},j-1)$ & $1$\\
\bottomrule
\end{tabular}
\label{tab:SR}
\end{table}

\vspace{-1em}
\subsection{Repetition Sequence}

In this subsection, a set of sequences, called repetition sequences, is defined that can be used to calculate the output bit estimates of an SR node based on the output bit estimates of its source node. To derive the repetition sequences, $\boldsymbol{v}$ is used to generate all the possible sequences that have to be XORed with the output of the source node to generate the output bit estimates of the SR node. Let $\eta_k$ denote the rightmost bit value of the left child node of the parent node of $\mathcal{N}^E_r$ at level $k+1$. When $v[k+1] = 0$, the left child node is a Rate-0 node so $\eta_k = 0$. When $v[k+1] = 1$, the left child node is a REP node, thus $\eta_k$ can take the value of either $0$ or $1$. The number of repetition sequences is dependent on the number of different values that $\eta_k$ can take. Let $W_{\boldsymbol{v}}$ denote the number of `$1$'s in $\boldsymbol{v}$. The number of all possible repetition sequences is thus $2^{W_{\boldsymbol{v}}}$. Let $\mathbb{S} = \{\boldsymbol{s}_1,\ldots,\boldsymbol{s}_{2^{W_{\boldsymbol{v}}}}\}$ denote the set of all possible repetition sequences.

The output bits of SR node $\beta_i^j[1:2^j]$ have the property that their repetition sequence is repeated in blocks of length $2^{j-r}$. Let $\beta_r^E[1:2^r]$ denote the output bits of the source node of an SR node $\mathcal{N}_j^i$. The output bits for each block of length $2^{j-r}$ in $\mathcal{N}_j^i$ with respect to the output bits of its source node can be written as
\begin{equation}
\label{eq:betaS}
\beta_j^i\left[\left(k-1\right)2^{j-r}+1:k2^{j-r}\right]=\beta_r^E\left[k\right]\oplus{\boldsymbol{s}_l},
\end{equation}
where $k\in\left\{1,\dots,2^r\right\}$ and $\boldsymbol{s}_l = \{s_l[1],\ldots,s_l[2^{j-r}]\}$ is the $l$-th repetition sequence in $\mathbb{S}$. To obtain the repetition sequence $\boldsymbol{s}_l$ and with a slight abuse of terminology and notation for convenience, the Kronecker sum operator $\boxplus$ is used, which is equivalent to the Kronecker product operator, except that addition in GF($2$) is used instead of multiplication. For each set of values that $\eta_k$'s can take, $\boldsymbol{s}_l$ can be calculated as
\begin{equation}
\boldsymbol{s}_l=\left(\eta_r,0\right)\boxplus\left(\eta_{r+1},0\right)\boxplus\cdots\boxplus \left(\eta_{j-1},0\right).
\label{eq:S}
\end{equation}

\begin{example}[Repetition sequences for $\mathrm{SR}((1,1),\text{EG-PC},2)$]
Consider an example in which the SR node $\mathcal{N}_4^1$ is located at level $4$ of the decoding tree and its source node $\mathcal{N}_2^4$ is an EG-PC node located at level $2$. Since $\boldsymbol{v} = (1,1)$, $W_{\boldsymbol{v}} = 2$ and $|\mathbb{S}| = 4$. For $\eta_1 \in \{0,1\}$ and $\eta_2 \in \{0,1\}$,
\begin{align*}
    \boldsymbol{s}_1 &= (0,0) \boxplus (0,0) = (0,0,0,0),\\ 
    \boldsymbol{s}_2 &= (1,0) \boxplus (0,0) = (1,1,0,0),\\
    \boldsymbol{s}_3 &= (0,0) \boxplus (1,0) = (1,0,1,0),\\ 
    \boldsymbol{s}_4 &= (1,0) \boxplus (1,0) = (0,1,1,0).
\end{align*}
\end{example}

For a polar code with a given $\boldsymbol{d}$, the locations of SR nodes in the decoding tree are fixed and can be determined off-line. Therefore, the repetition sequences in ${\mathbb{S}}$ of all of the SR nodes can be pre-computed and used in the course of decoding.

\vspace*{-1em}

\subsection{Decoding of SR Nodes}

To decode SR nodes, the LLR values $\alpha_{r_l}^E[1:2^r]$ of the source node $\mathcal{N}_r^E$ are calculated based on the LLR values $\alpha_{j}^i[1:2^j]$ of the SR node $\mathcal{N}_j^i$ for every repetition sequence $\boldsymbol{s}_l$ as follows:
\begin{proposition}
\label{pr:SN}
Let $\alpha_{j}^i[1:2^j]$ be the LLR values of the SR node $\mathcal{N}_j^i$ and $\alpha_{r_l}^E[1:2^r]$ be the LLR values of its source node $\mathcal{N}_r^E$ associated with the $l$-th repetition sequence $\boldsymbol{s}_l$. For $k\in\left\{1,\dots,2^r\right\}$ and $l\in\left\{1,\dots,2^{W_{\boldsymbol{v}}}\right\}$,
    \begin{equation}
\label{eq:alphaS}    
\alpha_{r_l}^E\left[k\right]= \sum_{m = 1}^{2^{j-r}} \alpha_{j}^i\left[\left(k-1\right)2^{j-r}+m\right] \left(-1\right)^{s_l[m]}.
    \end{equation}
\end{proposition} 
\begin{IEEEproof}   
See Appendix~\ref{app:a}.
\end{IEEEproof}

Using (\ref{eq:betaS}) and (\ref{eq:alphaS}), (\ref{eq:estimate}) can be written as
\begin{align}
\label{eq:estimate1}
\hat\beta_j^i\left[1:2^j\right] & = \underset{\beta_j^i\left[1:2^j\right]\in\mathbb{C}_j^i}{\argmax}\sum_{k=1}^{2^j}\left(-1\right)^{\beta_j^i\left[k\right]}\alpha_j^i\left[k\right]\\\nonumber
&=\mkern-10mu\underset{\substack{\beta_r^E\left[1:2^r\right]\in\mathbb{C}_r^E\\\boldsymbol s_l\in {\mathbb S}}}{\argmax}\!\sum_{k=1}^{2^r}\!\left(-\!1\right)^{\beta_r^E\left[k\right]}\!\sum_{m = 1}^{2^{j-r}}\!\!\alpha_{j_l}^i\!\!\left[\left(k\!-\!\!1\right)\!2^{j-r}\!\!+\!\!m\right]\!\left(-\!1\right)^{s_l[m]}\nonumber\\
&=\mkern-10mu\underset{\substack{\beta_r^E\left[1:2^r\right]\in\mathbb{C}_r^E\\l\in \{1,\ldots,\left|{\mathbb S}\right|\}}}{\argmax}\sum_{k=1}^{2^r}\left(-1\right)^{\beta_r^E\left[k\right]}\alpha_{r_l}^E\left[k\right].
\end{align}
Thus, the bit estimates of an SR node $\hat\beta_j^i\left[1:2^j\right]$ can be calculated by finding the bit estimates of its source node $\beta_r^E\left[1:2^r\right]$ using (\ref{eq:estimate1}) and the repetition sequences as shown in (\ref{eq:betaS}).

The decoding algorithm of an SR node $\mathcal{N}_j^i$ is described in Algorithm~\ref{alg:alg1}. It first computes $\alpha_{r_l}^E$ for $l\in\left\{1,\dots,\left|{\mathbb S}\right|\right\}$ and generates $\left|{\mathbb S}\right|$ new paths by extending the decoding path at the $j-r$ rightmost bits corresponding to $\eta_r,\eta_{r+1},\dots,\eta_{j-1}$. Note that the $l$-th path is generated when the repetition sequence is ${\boldsymbol s}_l$ and $\alpha_{r_l}^E$, $\widehat\beta_{r_l}^E$, and $\widehat \beta_{j_l}^i$ are its soft and hard messages. Then, the source node is decoded under the rule of the SC decoding. If the source node is a special node, a hard decision is made directly. Parity check and bit flipping will be performed further using Wagner decoder if the source node is an EG-PC node. Finally, the optimal decoding path index can be selected according to the comparison in (\ref{eq:comparison}) below and the decoding result is obtained using~(\ref{eq:betaS}).

Based on Algorithm~\ref{alg:alg1}, the SR node-based fast SC (SRFSC) decoding algorithm is proposed. It follows the SC decoding algorithm schedule until an SR node is encountered where Algorithm~\ref{alg:alg1} is executed. Note that the $\left|{\mathbb S}\right|$ paths can be processed simultaneously and the path selection operation in step 3 of Algorithm~\ref{alg:alg1} can be performed in parallel with the decoding of the source node in step 2 and the following $g$ function operation. Once the selected index $\hat l$ is obtained, only the $l$-th decoding path corresponding to ${\boldsymbol s}_l$ is retained and the remaining paths are deleted.


\begin{algorithm}
\small
\caption{Decoding algorithm of SR node $\mathcal{N}_j^i$} 
\begin{algorithmic}
\REQUIRE $\alpha_j^i\left[1:2^j\right]$, ${\mathbb S}$
\ENSURE $\hat\beta_j^i\left[1:2^j\right]$
\STATE 1) Soft message computation\\
\For{$l\in\left\{1,\dots,\left|{\mathbb S}\right|\right\}$}
{\STATE Calculate $\;\alpha_{r_l}^E$ according to (\ref{eq:alphaS}).
}
\vspace{1em}
\STATE 2) Decoding of source node $\mathcal{N}_r^E$\\
\For{$l\in\left\{1,\dots,\left|{\mathbb S}\right|\right\}$}{

\eIf{\text{SNT=Rate-C}}{Decode source node $\mathcal{N}_r^E$ using $\;\alpha_{r_l}^E$ and obtain $\hat\beta_{r_l}^E$\;}
{\eIf{\text{SNT=Rate-0}}{$\hat\beta_{r_l}^E\left[k\right]=0,\;k\in\left\{1,\dots,2^r\right\}$\;
}
(\tcp*[h]{\text{SNT=Rate-1 or SNT=EG-PC}})
{$\hat\beta_{r_l}^E\left[k\right]=h\left(\alpha_{r_l}^E\left[k\right]\right),\;k\in\left\{1,\dots,2^r\right\}$.}
}
\vspace{1em}
\If{\text{SNT=EG-PC}}{Perform parity check and bit flipping on $\hat\beta_{r_l}^E$ using $\alpha_{r_l}^E$.}
}

\vspace{1em}
\STATE 3) Comparison and path selection
\begin{equation}
\label{eq:comparison}
\hat l = \underset{l\in\left\{1,\dots,\left|\mathbb S\right|\right\}}{\argmax}\sum_{k=1}^{2^r}\left|\alpha_{r_l}^E\left[k\right]\right|.
\end{equation}
Return $\hat \beta_{j_{\hat l}}^i$ to parent node according to (\ref{eq:betaS}).

\end{algorithmic}
\label{alg:alg1}
\end{algorithm}


\section{Hard-decision-aided Fast SC Decoding with Sequence Repetition Nodes}
\label{sec:TA} 
In this section, a novel threshold-based hard-decision-aided scheme is proposed, which speeds up the decoding of general nodes with no specific structure in the SRFSC decoding at high signal-to-noise ratios. Consequently, the \emph{threshold-based hard-decision-aided SRFSC} (TA-SRFSC) decoding algorithm is proposed. In addition, a multi-stage decoding strategy is introduced to mitigate the possible error-correction performance loss of the proposed TA-SRFSC decoding.

\subsection{Proposed Threshold-based Hard-decision-aided Scheme}

In a binary AWGN channel, the LLR value $\alpha_j^i\left[k\right]$ of any node $\mathcal{N}_j^i$ can be approximated as a Gaussian variable as shown in Fig.~\ref{fig:distribution}. 
The red area in Fig.~\ref{fig:distribution} represents the approximate probability of correct hard decision $\widetilde P_c$ and the blue area represents the approximate probability of incorrect hard decision $\widetilde P_e$ when $\beta_j^i\left[k\right]=0$ such that
\begin{equation}
\widetilde P_c=Q\left(\frac{T-m_j^i}{\sqrt{2m_j^i}}\right),\;\;\widetilde P_e=Q\left(\frac{T+m_j^i}{\sqrt{2m_j^i}}\right),\label{eq:PcPe}
\end{equation}
where $Q\left(x\right)=\frac1{\sqrt{2\mathrm\pi}}\int_x^\infty e^{-\frac{t^2}2}dt$. The area between the two dashed lines represents the approximate probability that a hard decision is not performed.

\begin{figure}[t]
\centering
{\includegraphics{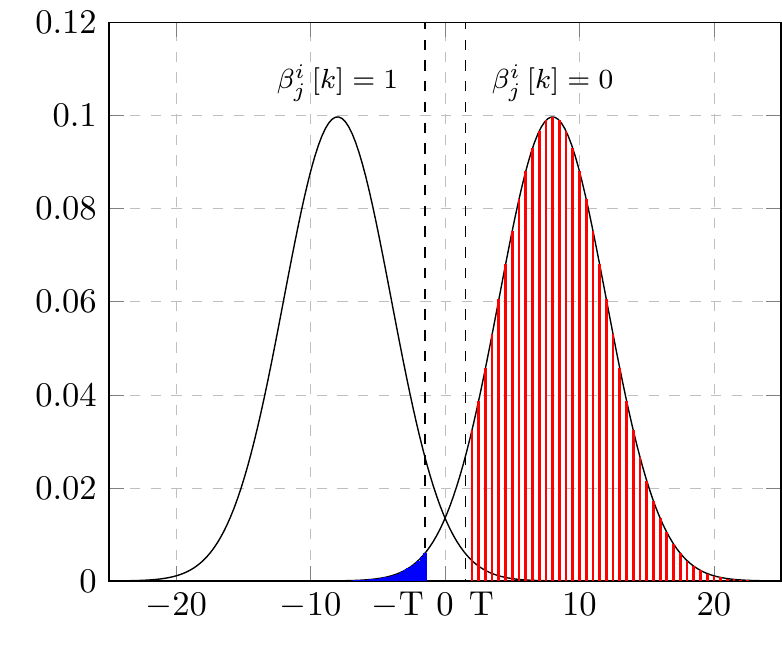}}
\setlength{\abovecaptionskip}{-5pt}
\caption{Probability distribution of $\alpha_j^i\left[k\right]$ under Gaussian approximation for $m_j^i=8$. The red dashed area represents the probability of correct hard decision and the blue solid area represents the probability of incorrect hard decision when $\beta_j^i\left[k\right]=0$.}
\label{fig:distribution}
\setlength{\belowcaptionskip}{-10pt}
\vspace{-1em}
\end{figure}

To simplify the calculation of the threshold, we take a different approach than \cite{li2018low} by using the Gaussian distribution of $\alpha_j^i\left[k\right]$ and constraining the approximate probability of error when $\beta_j^i\left[k\right]=0$ to be
\begin{equation}\label{eq:error constrain}
\widetilde  P_e=Q\left(\frac{T+m_j^i}{\sqrt{2m_j^i}}\right)<Q\left(c\right),
\end{equation}
where $c$ (and thus $Q\left(c\right)$) is a positive constant, whose selection method will be given in Observation~\ref{pr:LB}. This is equivalent to $T>-m_j^i+c\sqrt{2m_j^i}$. 
Therefore, the threshold is
\begin{equation} \label{eq:threshold}
   T=\left|-m_j^i+c\sqrt{2m_j^i}\right|,
\end{equation}
where the absolute value ensures $T$ is positive for all values of $m_j^i>0$ with any $c>0$. 

Under the Gaussian approximation in \eqref{eq:gaussian}, for a node that undergoes hard decision in (\ref{eq:hard decision}), the proposed threshold leads to a bounded probability of correct hard decision as shown in the following observation. Note that the analytical derivations are accurate under the assumption of Gaussian approximation.
\begin{observation}
\label{pr:LB}
Let $0.5<\varepsilon<1$ and $c>0$ be real numbers such that
\begin{equation}     \label{eq:ineq6} 
Q\left(c\right) \leq 1-\sqrt[2^n]{\varepsilon}.
\end{equation}
Performing hard decision in (\ref{eq:hard decision}) with the threshold in (\ref{eq:threshold}) on nodes $\mathcal{N}_j^i$ whose $m_j^i$ satisfy
\begin{equation} \label{eq:ineq11}
m_j^i\geq\frac{1}{2}\left[c-Q^{-1}\left(\frac{Q\left(c\right)}{\frac{1}{\sqrt[2^n]{\varepsilon}}-1}\right)\right]^2,
\end{equation}
has a probability of correct hard decision that is lower bounded by $\sqrt[2^{\left(n-j\right)}]\varepsilon$, under the assumption of Gaussian approximation and assuming all prior bits are decoded correctly.
\end{observation}
\textit{Explanation:} See Appendix~\ref{app:b}.

The proposed TA scheme performs hard decision on a node $\mathcal{N}_j^i$ only if (\ref{eq:ineq11}) is satisfied. Furthermore, a hard decision is performed on a node if all of its input LLR values $\alpha_j^i\left[k\right]$ satisfy (\ref{eq:hard decision}). Otherwise, standard SC decoding is applied on $\mathcal{N}_j^i$ to obtain the decoding result. Compared to the TA scheme in \cite{li2018low}, the threshold values of both methods can be computed off-line. However, the proposed TA scheme in this paper has the following three advantages: 1) The proposed method can provide a higher latency reduction than the best case in \cite{li2018low} as shown in the simulation results; 2) The effect of the proposed threshold values on the error-correction performance is approximately predictable according to Observation~\ref{pr:LB}; 3) According to \eqref{eq:ineq11}, only a fraction of nodes undergo hard decision in the decoding process of the proposed TA scheme, which avoids unnecessary threshold comparisons. To speed up the decoding process, the proposed TA scheme is combined with SRFSC decoding that results in the TA-SRFSC decoding algorithm. In TA-SRFSC decoding, when one of the special nodes considered in SRFSC decoding is encountered, SRFSC decoding is performed and when a general node with no special structure is encountered, the proposed TA scheme is applied. The following observation provides an approximate upper bound on the BLER of the proposed TA-SRFSC decoding.

\begin{observation}
\label{pr:BLER}
Let ${\mathrm{BLER}}_{\mathrm{TA-SRFSC}}$ and ${\mathrm{BLER}}_{\mathrm{SRFSC}}$ denote the BLER of the TA-SRFSC decoding and the SRFSC decoding respectively. We can have the following approximate inequality
\begin{equation}
\label{eq:performance5}
{\mathrm{BLER}}_{\mathrm{TA-SRFSC}}\lesssim 1-\varepsilon\left(1-{\mathrm{BLER}}_{\mathrm{SRFSC}}\right).
\end{equation} 
\end{observation}
\textit{Explanation:} See Appendix~\ref{app:c}.

Observation~\ref{pr:BLER} provides a method to derive the threshold value for a desired upper bound of the BLER for TA-SRFSC decoding approximately. In fact, a large threshold value results in a better error-correction performance than a small threshold value at the cost of lower decoding speed. 
Therefore, a trade-off between the error-correction performance and the decoding speed can be achieved with the proposed TA-SRFSC decoding algorithm.


\subsection{Multi-stage Decoding}
\label{sec:multistage}

To mitigate the possible error-correction performance loss of the proposed
TA-SRFSC decoding, a multi-stage decoding strategy is adopted in which a maximum of two decoding attempts is conducted. In the first decoding attempt, TA-SRFSC decoding is used. If this decoding fails and if there existed a node that underwent hard decision in this first decoding attempt, then a second decoding attempt using SRFSC decoding is conducted. To determine if the TA-SRFSC decoding failed, a cyclic redundancy check (CRC) is concatenated to the polar code and it is verified after the TA-SRFSC decoding. Additionally, the CRC is verified after the second decoding attempt by the SRFSC decoding to determine if the overall decoding process succeeded.

As shown in the next section, most of the received frames are decoded correctly by the proposed TA-SRFSC decoding in the first decoding attempt. As a result, the average decoding latency of the proposed multi-stage SRFSC decoding is very close to that of the TA-SRFSC decoding with CRC bits. However, the error-correction performance of the proposed multi-stage SRFSC decoding algorithm is slightly worse than that of SRFSC decoding due to the addition of CRC bits. As such, multi-stage SRFSC decoding trades off a slight degradation in error-correction performance to obtain a significant reduction in the average decoding latency. It is worth noting that, in order to improve the error-correction performance of the proposed scheme, the proposed multi-stage decoding strategy can be generalized to have more than two decoding attempts. In this scenario the first two attempts are the same as described above, i.e., TA-SRFSC decoding is used first followed by SRFSC decoding.\footnote{If the TA-SRFSC decoder does not satisfy the CRC, the SRFSC decoding either should start from the beginning or from the first node which underwent the hard decision. In this paper, the SRFSC decoder starts decoding from the first bit to avoid storing intermediate LLRs.} The third and any subsequent attempts would use increasingly more powerful decoding techniques than the first two, such as SCL decoding.

\section{Decoding Latency}
\label{sec:performance}

In this section, the decoding latency of the proposed fast decoders is analyzed under two cases: 1) with no resource limitation and 2) with hardware resource constraints. The former is to facilitate the comparisons with the works that do not consider hardware implementation constraints \cite{hanif2017fast,condo2018generalized}, and the latter is for comparison with those that do \cite{sarkis2014fast,giard2018fast,ercan2019operation}.
\vspace{-1em}

\subsection{No Resource Limitation}
\label{sec:nolimitation}
In this case, the same assumptions as in \cite{arikan2009,hanif2017fast} are used and the decoding latency is measured in terms of the number of required time steps. More specifically: 1) there is no resource limitation so that all the parallelizable instructions are performed in one time step, 2) bit operations are carried out instantaneously, 3) addition/subtraction of real numbers and check-node operation consume one time step, and 4) Wagner decoding can be performed in one time step.

\subsubsection{SRFSC and TA-SRFSC}
\label{sec:SRFSC}

For any node $\mathcal{N}_j^i$ that has no special structure, e.g. a parent node of an SR node, and that satisfies (\ref{eq:ineq11}), the threshold comparison in (\ref{eq:hard decision}) is performed in parallel with the calculation of the LLR values of its left child node. Therefore, the proposed hard decision scheme does not affect the latency requirements for the nodes that undergo hard decision. 

The number of time steps required for the decoding of the SR node is calculated according to Algorithm~\ref{alg:alg1}. In Step 1, the calculation of the LLR values for the source node requires one time step if $\boldsymbol{v} \neq \emptyset$. If $\boldsymbol{v} = \emptyset$, then the LLR values of the source node are available immediately. Thus, the required number of time steps for Step 1 is
\begin{equation}
{\mathcal{T}}_1=\begin{cases} 0, & \text{if }\boldsymbol{v} = \emptyset,\\1, & \text{if }\boldsymbol{v} \neq \emptyset.
\end{cases}
\end{equation}

The time step requirement of Step 2 depends on the source node type. If $\text{SNT}=\text{Rate-C}$, the time step requirement of Step 2 is the time step requirement of the $\text{Rate-C}$ node. If $\text{SNT}=\text{Rate-0}$ or $\text{Rate-1}$, then there is no latency overhead in Step 2. If $\text{SNT}=\text{EG-PC}$ and in accordance with Section~\ref{sec:source node}, $z$ can be estimated in two time steps if the leftmost node is a REP node (one time step for performing the check-node operation and one time step for adding the LLR values). Also, Wagner decoding can be performed in parallel with the estimation of $z$ assuming $z=0$ or $z=1$. As such, at most two time steps are required for parity check and bit flipping of the $\text{EG-PC}$ node. The required number of time steps for Step 2 is
\begin{equation}
{\mathcal{T}}_2=\begin{cases} 0, & \text{if }\text{SNT}=\text{Rate-0 or Rate-1},\\1 \text{ or } 2, & \text{if }\text{SNT}=\text{EG-PC},\\2^{r+1}-2, & \text{if }\text{SNT}=\text{Rate-C}.
\end{cases}
\end{equation}

Step 3 consumes two time steps using an adder tree and a comparison tree if $\left|\mathbb S\right|>1$ and no time steps otherwise. Thus, the number of required time steps for Step 3 is
\begin{equation}
{\mathcal{T}}_3=\begin{cases} 0, & \text{if }\left|\mathbb S\right|=1,\\2, & \text{if }\left|\mathbb S\right|>1.
\end{cases} \end{equation}

Since path selection in Step 3 can be executed in parallel with the decoding of source node in Step 2 and the following $g$ function calculation, the total number of time steps required to decode an SR node can be expressed as
\begin{equation}
{\mathcal{T}}_\mathrm{SR}={\mathcal{T}}_1+\max\left({\mathcal{T}}_2,{\mathcal{T}}_3-1\right),
\end{equation}
where ${\mathcal{T}}_3-1$ indicates that at least one time step in ${\mathcal{T}}_3$ can be reduced by parallelizing Step 3 and the $g$ function calculation. Therefore, ${\mathcal{T}}_\mathrm{SR}$ is a variable that is dependent on its parameters. However, with a given polar code, the total number of time steps required for the decoding of the polar code using SRFSC decoding is fixed, regardless of the channel conditions.

\subsubsection{Multi-stage SRFSC}

Let ${\mathcal{T}}_{\mathrm{TA-SRFSC}_\mathrm{crc}}$ and ${\mathcal{T}}_{\mathrm{SRFSC}_\mathrm{crc}}$ denote the average decoding latency of the proposed TA-SRFSC decoding and the SRFSC decoding using CRC bits, respectively. The average decoding latency of multi-stage SRFSC decoding, ${\mathcal{T}}_{\mathrm{Multi-stage\;SRFSC}}$, is
\begin{equation}\label{eq:latency1}
\mathcal{T}_{\mathrm{Multi-stage\;SRFSC}}={\mathcal{T}}_{\mathrm{TA-SRFSC}_\mathrm{crc}}+P_{\mathrm{Re-decoding}}{\mathcal{T}}_{\mathrm{SRFSC}_\mathrm{crc}},
\end{equation}
where $P_{\mathrm{Re-decoding}}$ indicates the probability that TA-SRFSC decoding fails and there is at least one node that undergoes hard decision. Note that $P_{\mathrm{Re-decoding}}$ is less than or equal to the probability that the output of TA-SRFSC decoding fails the CRC verification, which can be approximated by ${\mathrm{BLER}}_{\mathrm{TA-SRFSC}_\mathrm{crc}}$. The approximation is due to the fact that the undetected error probability of CRC is negligible when its length is long enough \cite{el2015detection}. In accordance with Observation~\ref{pr:BLER}, the average decoding latency requirement for the proposed multi-stage SRFSC decoding is
\begin{equation}\label{eq:latency2}
\begin{array}{l}\mathcal{T}_{\mathrm{Multi-stage\;SRFSC}}\\ \lesssim {\mathcal{T}}_{\mathrm{TA-SRFSC}_\mathrm{crc}}+\left(1-\varepsilon\left(1-{\mathrm{BLER}}_{\mathrm{SRFSC}_\mathrm{crc}}\right)\right){\mathcal{T}}_{\mathrm{SRFSC}_\mathrm{crc}}.\end{array}
\end{equation}

Since the decoding latency of SRFSC decoding is fixed, the average decoding latency and the worst case decoding latency of SRFSC decoding are equivalent. The worst case decoding latency of the proposed TA-SRFSC decoding can be calculated when none of the nodes in the decoding tree undergo hard decision. This occurs when the channel has a high level of noise. Thus, the worst case decoding latency of the proposed TA-SRFSC decoding is equivalent to the decoding latency of the SRFSC decoding. Moreover, when the channel is too noisy, $P_{\mathrm{Re-decoding}}\approx 0$, because almost none of the nodes undergo hard decision. Thus, the worst case decoding latency of the proposed multi-stage SRFSC decoding is equivalent to the worst case decoding latency of TA-SRFSC decoding using a CRC, which is the latency of SRFSC decoding with a CRC.

\subsection{With Hardware Resource Constraints}
\label{sec:limitation}
In this case, the decoding latency is measured in terms of the number of clock cycles in the actual hardware implementation. Contrary to Section~\ref{sec:nolimitation}, a time step may take more than one clock cycle in a realistic hardware implementation. First, the idea of a semi-parallel decoder in \cite{leroux2012semi} is adopted in which at most $P$ processing elements can run at the same time. Thus, parallel $f$ function or $g$ function operations on more than $P$ processing elements require more than one clock cycle. Second, the adder tree used for the calculation of LLR values in step 1 of Algorithm~\ref{alg:alg1}, and the compare-and-select (CS) tree used by the Wagner decoder to find the index of the least reliable input bit in step 2 of Algorithm~\ref{alg:alg1}, both require pipeline registers to reduce the critical path delay and improve the operating frequency and the overall throughput. The addition of these pipeline registers increases the number of decoding clock cycles. The exact number depends on the implementation and the given polar code.

\section{Results and Comparison}
\label{sec:Numerical results}

In this section, the average decoding latency and the error-correction performance of the proposed decoding algorithms are analyzed and compared with state-of-the-art fast SC decoding algorithms. 
To derive the results, polar codes of length $N\in\left\{128,512,1024\right\}$, which are adopted in the 5G standard \cite{3GPP1}, are used. To do a fair comparison, for baseline decoding algorithms without hardware implementation, the latency is calculated under the assumptions in Section~\ref{sec:nolimitation}. Otherwise, the hardware implementation results are reported.

To simulate the effect of $\varepsilon$ on the error-correction performance and the latency of the proposed decoding algorithms, the assumption in \cite{li2018low} is adopted and three values of $\varepsilon\in\{0.9,0.99,0.999\}$ are selected. In accordance with (\ref{eq:ineq6}), $c\geq3.8$ for $\varepsilon=0.9$, $c\geq4.3$ for $\varepsilon=0.99$, and $c\geq4.8$ for $\varepsilon=0.999$. According to (\ref{eq:error constrain}), with the increasing of $c$, $\widetilde P_e$ decreases. At the same time, fewer nodes undergo hard decision and the decoding latency increases. To get a good trade-off between error-correction performance and latency, we set $c=3.8$ for $\varepsilon=0.9$, $c=4.3$ for $\varepsilon=0.99$, and $c=4.8$ for $\varepsilon=0.999$. Consequently, $m_j^i\geq9.3891$ for $\varepsilon=0.9$, $m_j^i\geq14.7255$ for $\varepsilon=0.99$, and $m_j^i\geq16.1604$ for $\varepsilon=0.999$ in accordance with (\ref{eq:ineq11}). Using these values, the threshold $T$ defined in (\ref{eq:threshold}) and the BLER upper bound for the TA-SRFSC decoding in (\ref{eq:performance5}) can be calculated for different values of $\varepsilon$.

\begin{table}[htbp] 
\begin{minipage}[t]{0.5\textwidth}
\centering
\caption{The number of SR nodes with different $\left|{\mathbb S}\right|$ and the number of general nodes in 5G polar codes of lengths $N \in \{128,512,1024\}$ and rates $R \in \{1/4,1/2,3/4\}$.}
\setlength{\belowcaptionskip}{-10pt}
\vspace{-1em}
\scalebox{1}{\renewcommand\arraystretch{0.5}
        \begin{tabular}{p{0.4cm}<{\centering}p{0.4cm}<{\centering}p{0.3cm}<{\centering}p{0.3cm}<{\centering}p{0.3cm}<{\centering}p{0.3cm}<{\centering}p{0.3cm}<{\centering}p{0.75cm}<{\centering}p{1cm}<{\centering}c}
            \toprule
            & & \multicolumn{6}{c}{SR} & General \\
            \cmidrule(lr){3-8} \cmidrule(lr){9-9}
            \multirow{2}{*}{$N$} & \multirow{2}{*}{$R$} & \multicolumn{5}{c}{$\left|{\mathbb S}\right|$} &\multirow{2}{*}{Total} & \multirow{2}{*}{Total} \\
            \cmidrule(lr){3-7}
            & & 1 & 2 & 4 & 8 & 16 & & & \\                        
            \midrule
            \multirow{3}{*}{128} & $1/4$ & 1 & 0 & 2 & 1 & 0 & 4 & 3\\
          & $1/2$ & 4 & 3 & 1 & 0 & 0 & 8 & 7\\  
          & $3/4$ & 8 & 2 & 0 & 0 & 0 & 10 & 9\\ 

            \midrule
            \multirow{3}{*}{512} & $1/4$ & 12 & 2 & 2 & 1 & 0 & 17 & 16 \\
          & $1/2$ & 15 & 5 & 2 & 0 & 1 & 23 & 22\\  
          & $3/4$ & 13 & 5 & 1 & 0 & 1 & 20 & 19\\                    
            \midrule
            \multirow{3}{*}{1024} & $1/4$ & 17 & 6 & 2 & 2 & 1 & 28 & 27 \\
          & $1/2$ & 25 & 8 & 2 & 3 & 1 & 39 & 38\\  
          & $3/4$ & 29 & 8 & 2 & 1 & 0 & 40 & 39\\                    
        \bottomrule
        \end{tabular}}
\label{tab:nodenumber}  
\end{minipage}
\hfill
\begin{minipage}[t]{0.5\textwidth}
\centering
\small
\caption{Node length of SR nodes with different $\left|{\mathbb S}\right|$ in 5G polar codes of lengths $N \in \{128,512,1024\}$ and rates $R=1/2$.}
\setlength{\belowcaptionskip}{-10pt}
\renewcommand\arraystretch{0.5}
        \begin{tabular}{p{0.4cm}<{\centering}p{0.8cm}<{\centering}p{0.25cm}<{\centering}p{0.25cm}<{\centering}p{0.25cm}<{\centering}p{0.25cm}<{\centering}p{0.25cm}<{\centering}c}
            \toprule
            \multirow{2}{*}{$N$} & \multirow{2}{*}{\rotatebox{0}{\makecell[c]{Length}}} & & & $\left|{\mathbb S}\right|$& & \\
            \cmidrule(lr){3-7}
            & & 1 & 2 & 4 & 8 & 16\\                                                 
            \midrule
            \multirow{3}{*}{128} & $8$ & 2 & 2 & 0 & 0 & 0 \\
          & $16$ & 0 & 1 & 1 & 0 & 0 \\ 
          & $32$ & 2 & 0 & 0 & 0 & 0 \\ \midrule     
            \multirow{5}{*}{512} & $8$ & 7 & 3 & 0 & 0 & 0\\
          & $16$ & 4 & 1 & 2 & 0 & 0 \\  
          & $32$ & 3 & 1 & 0 & 0 & 0 \\                    
          & $64$ & 1 & 0 & 0 & 0 & 0 \\ & $128$ & 0 & 0 & 0 & 0 & 1 \\                             
\midrule            
            \multirow{5}{*}{1024} & $8$ & 10 & 6 & 0 & 0 & 0 \\
          & $16$ & 7 & 1 & 2 & 0 & 0 \\  
          & $32$ & 4 & 0 & 0 & 3 & 0 \\                    
          & $64$ & 2 & 1 & 0 & 0 & 1 \\                              
          & $128$ & 2 & 0 & 0 & 0 & 0 \\                                        
\bottomrule
        \end{tabular}
\label{tab:nodelength}
\end{minipage} 
\end{table}

Table~\ref{tab:nodenumber} reports the number of SR nodes with different $\left|{\mathbb S}\right|$, the total number of SR nodes, and the total number of general nodes with no special structure, at different code lengths and rates. It can be seen that when the code length is $128$, $512$, and $1024$, the codes with rate $1/2$ have the largest proportion of nodes with $\left|\mathbb S\right|>1$, respectively. This in turn results in more latency savings because a higher degree of parallelism can be exploited with these nodes. Thanks to the binary tree structure and the fact that SR nodes are always present at an intermediate level of the decoding tree, the number of traversed general nodes with no special structure equals the total number of SR nodes minus one, since they are the parent nodes of SR nodes. Table~\ref{tab:nodelength} shows the length of SR nodes with different $\left|{\mathbb S}\right|$ at different code lengths when $R=1/2$. The length of the SR nodes in the decoding tree corresponds to the level in the decoding tree that they are located. SR nodes with larger $\left|{\mathbb S}\right|$ that are located on a higher level of the decoding tree contribute more in the overall latency reduction.

Table~\ref{tab:cycle} reports the number of time steps required to decode polar codes of lengths $N \in \{128,512,1024\}$ and rates $R \in \{1/4,1/2,3/4\}$ with the proposed SRFSC decoding algorithm, and compares it with the required number of time steps of the decoders in \cite{sarkis2014fast ,giard2018fast,hanif2017fast,condo2018generalized,ercan2019operation}. Note that the decoder in \cite{ercan2019operation} has the minimum required time steps amongst all the previous works considered in Table~\ref{tab:cycle}. Three versions of the proposed SRFSC decoder are considered in Table~\ref{tab:cycle}: the SRFSC decoder that only utilizes SR nodes; the SRFSC$^{'}$ decoder that considers SR nodes and P-0SPC nodes that is adopted in all other baseline decoders in Table~\ref{tab:cycle}; and the SRFSC$^{''}$ decoder that uses SR nodes, P-0SPC nodes, and the operation mergers $\text{F}^{\times2}$ and $\text{G-F}$. It can be seen that the SRFSC$^{''}$ decoder requires fewer time steps with respect to other decoders, except for the case of $N=128$ and $R=3/4$, which has a frequent occurrence of REP-SPC nodes that provides an advantage for \cite{ercan2019operation}. This is due to the fact that the REP-SPC decoder in other works decodes REP and SPC nodes in parallel, consuming only a single time step, while two time steps are needed to decode the REP-SPC node in the proposed SRFSC decoder.

\begin{table*}[t]
\centering
\caption{Number of time steps for different fast SC decoding algorithms of polar codes of lengths $N \in \left\{128,512,1024\right\}$ and rates $R \in \left\{1/4,1/2,3/4\right\}$.}
\setlength{\belowcaptionskip}{-10pt}
\label{tab:cycle}
  \renewcommand\arraystretch{0.5}
        \begin{tabular}{p{0.5cm}<{\centering}p{0.5cm}<{\centering}p{0.65cm}<{\centering}p{0.65cm}<{\centering}p{0.65cm}<{\centering}p{0.65cm}<{\centering}p{0.65cm}<{\centering}p{0.75cm}<{\centering}p{0.75cm}<{\centering}c}
            \toprule
            $N$ & $R$ &  \cite{sarkis2014fast} & \cite{giard2018fast} &\cite{hanif2017fast} & \cite{condo2018generalized} & \cite{ercan2019operation} & SRFSC & SRFSC$^{'}$ & SRFSC$^{''}$\\
            \midrule
            \multirow{3}{*}{128} & $1/4$ & 25 & 23 & 24 & 23 & 10 &13 &12&10 \\
            &  $1/2$ & 33 & 31 &24 & 24 &  21 & 25&24&20 \\
            & $3/4$ & 22 & 22 &22 & 22 &  16 & 29&25&20 \\
\midrule
            \multirow{3}{*}{512} & $1/4$ & 73 & 70 & 63 & 63 & 42 & 57&52&40 \\
            &  $1/2$ &  87 & 83 &77 & 77 & 56 &72& 66&52 \\
            & $3/4$ &  79 & 73 & 64 & 64 &50 & 63&56&45 \\
\midrule
            \multirow{3}{*}{1024} & $1/4$ &  122 & 115 & 110 & 108 &68 & 92&85&66 \\
            &  $1/2$ &  156 & 144 &138 & 138 & 94 & 127&115 &90\\
            & $3/4$ &  138 & 132 &116 & 116 & 89 & 123&111&86 \\
\bottomrule
        \end{tabular}
\vspace{-2em}
\end{table*}

Table~\ref{tab:nodetypenumber} compares the required number of clock cycles and the maximum operating frequency in the hardware implementation of different decoders for a polar code with $N=1024$, $R \in \{1/2,1/4,3/4\}$, and $P=64$ for the proposed SRFSC decoding algorithm (taken from \cite{zheng2020implementation}) and the decoders in \cite{sarkis2014fast,ercan2017reduced,ercan2019operation,giard2018fast}. All FPGA results are for an Altera Stratix IV EP4SGX530KH40C2 FPGA device and all ASIC results use TSMC 65nm CMOS technology. The required number of clock cycles for the proposed SRFSC decoder can be further reduced if we consider node-branch merging and branch operation merging similar to \cite{ercan2019operation}. We have verified that the introduction of P-RSPC node to the proposed SRFSC decoder lengthens the critical path of the decoder, but the merging of branch operations, $\text{F}^{\times2}$ and $\text{G-F}$, does not have an effect on the operating frequency. Thus, an enhanced SRFSC decoder with merging branch operations $\text{F}^{\times2}$ and $\text{G-F}$ is implemented and denoted as SRFSC$^\ast$ in Table~\ref{tab:nodetypenumber}. It can be seen that the SRFSC$^\ast$ decoder requires a smaller number of clock cycles compared to other decoders and the reduction with respect to \cite{ercan2019operation} is $8\%$, $4\%$, and $8\%$ at $R\in\{1/4,1/2,3/4\}$, respectively. Moreover, the SRFSC decoder and the SRFSC$^\ast$ decoder both have the highest maximum operating frequency $f_{\max}$ when implemented on the FPGA. Compared with the works that reported $f_{\max}$ results for FPGA implementations, our decoders achieve an improvement of at least $10\%$.

\begin{table}[t]
\centering
\caption{Required number of clock cycles and maximum operating frequency for different decoding algorithms of polar codes of length $N=1024$ with rate $R=\left\{1/4,1/2,3/4\right\}$ and $P=64$.}
\label{tab:nodetypenumber}
 \renewcommand\arraystretch{0.8}

        \begin{tabular}{p{0.8cm}<{\centering}p{0.6cm}<{\centering}p{0.6cm}<{\centering}p{0.6cm}<{\centering}p{0.6cm}<{\centering}p{0.6cm}<{\centering}c}

            \toprule
            & \multicolumn{3}{c}{clock cycles} & \multicolumn{2}{c}{$f_{\max} (\text{MHz})$}\\
            \cmidrule(lr){2-4}\cmidrule(lr){5-6}
            
            \multirow{2}{*}{} & \multicolumn{3}{c}{$R$} &\multirow{2}{*}{FPGA} & \multirow{2}{*}{ASIC}\\
            \cmidrule(lr){2-4}
            & $1/4$ & $1/2$ & $3/4$ & &\\                        
            \midrule
            $\text{SRFSC}$ & 186 & 222 & 200 & 109.6 & \textemdash\\
            ${\text{SRFSC}}^\ast$ & 155 & 191 & 166 & 109.6 & \textemdash\\
            \cite{ercan2019operation} & 168 & 198 & 181 & \textemdash & 430\\          
            \cite{ercan2017reduced} & 189 & 214 & 185 & 89.6 & 420\\ 
            \cite{giard2018fast} & 221 & 252 & 216 & \textemdash & 450\\
            \cite{sarkis2014fast} & 225 & 270 & 225 & 99.8 & 450\\
        \bottomrule
        \end{tabular}
\end{table}

\begin{figure}[t]
\centering
\scalebox{0.75}{\includegraphics{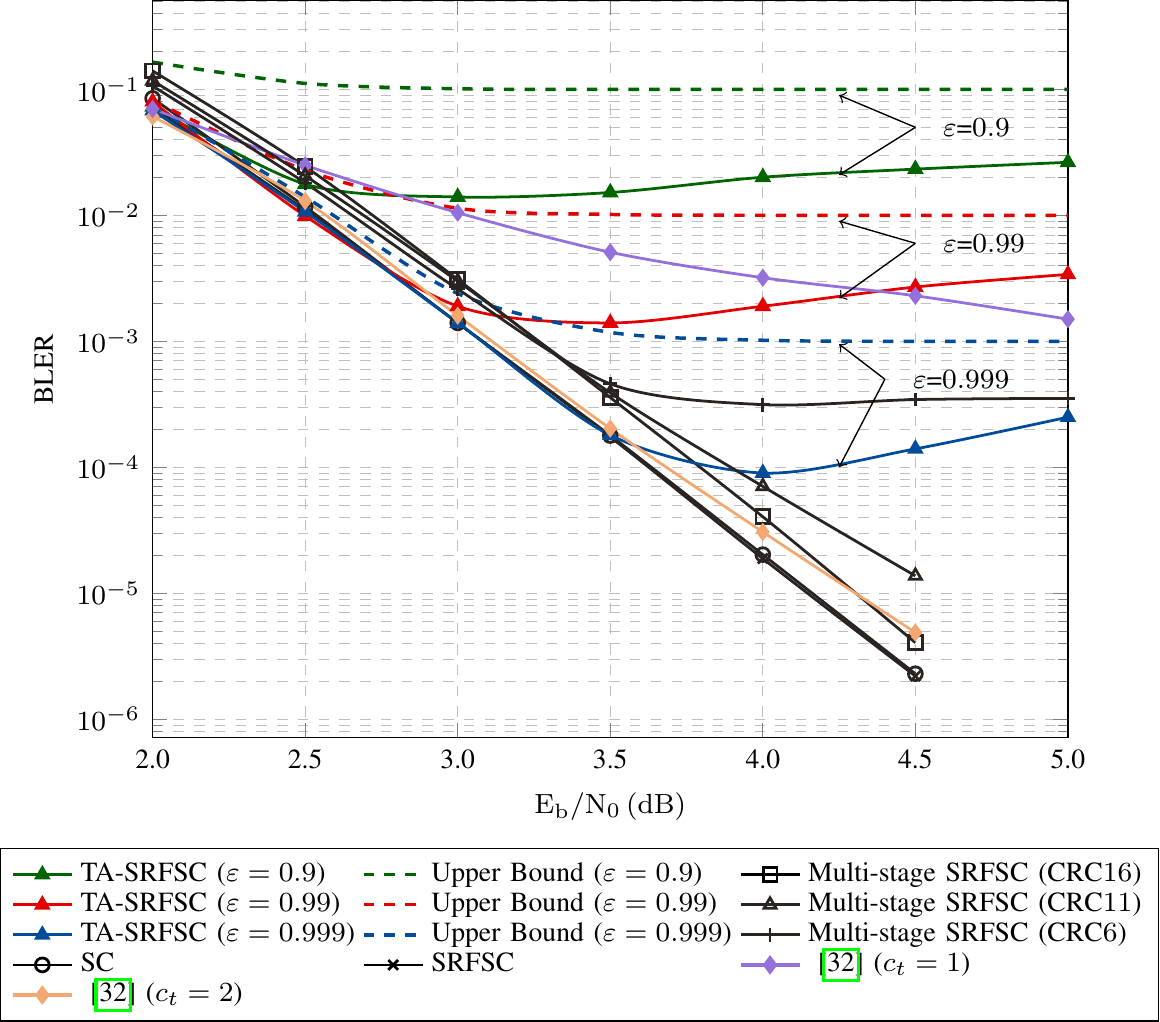}}
\setlength{\abovecaptionskip}{-5pt}
\caption{BLER performance of different decoding algorithms for the 5G polar code of length $N=1024$ and rate $R=1/2$.} 
  \label{fig:BLER}
\end{figure}

Fig.~\ref{fig:BLER} shows the BLER performance of different decoding algorithms when $N=1024$ and $R=1/2$, for different values of energy per bit to noise power spectral density ratio ($\mathrm{E_b}/\mathrm N_0$). For each value of $\varepsilon$, the BLER of TA-SRFSC decoding is depicted together with the upper bound calculated in Observation~\ref{pr:BLER}. It can be seen that the introduction of the TA scheme results in BLER performance loss for the proposed TA-SRFSC decoding with respect to SC and SRFSC decoding, especially at higher values of $\mathrm{E_b}/\mathrm{N_0}$. The simulations confirm that the BLER curves of TA-SRFSC decoding fall below their respective upper bounds. It can also be seen that, as the $\mathrm{E_b}/\mathrm{N_0}$ value increases beyond a specific point, the BLER performance of the TA-SRFSC decoding degrades. This is because of the difference in the performance of hard-decision and soft-decision decoding. In accordance with (\ref{eq:ineq11}), more nodes undergo hard decision decoding for larger values of $\mathrm{E_b}/\mathrm{N_0}$. Therefore, while the channel conditions improve, the hard decision decoding introduces errors that reduce the error-correction performance gain associated with these large $\mathrm{E_b}/\mathrm{N_0}$ values. As a result, the BLER performance degrades after a certain value of $\mathrm{E_b}/\mathrm{N_0}$. This phenomenon exists as long as there are nodes that can undergo hard decision decoding. After all the nodes are decoded using hard decision, the BLER performance improves again as $\mathrm{E_b}/\mathrm{N_0}$ increases.

The proposed multi-stage SRFSC decoder is implemented using three different CRC lengths that are adopted in 5G to identify whether the decoding succeeded or failed: the CRC of length $6$ (CRC$6$) with generator polynomial $D^6+D^5+1$, the CRC of length $11$ (CRC$11$) with generator polynomial $D^{11}+D^{10}+D^9+D^5+1$, and the CRC of length $16$ (CRC$16$) with generator polynomial $D^{16}+D^{12}+D^5+1$. For all values of $\varepsilon$, the proposed multi-stage SRFSC decoding results in almost the same BLER performance. Therefore, only the curve with $\varepsilon=0.9$ is plotted in Fig.~\ref{fig:BLER} for the multi-stage SRFSC decoding. CRC$16$ provides a better error-correction performance compared to CRC$6$ and CRC$11$, especially at high values of $\mathrm{E_b}/\mathrm{N_0}$, due to high undetected probability of error for short CRC lengths. Hence, CRC$16$ is selected for the polar code of length $N=1024$ in this paper. It can be seen that the multi-stage SRFSC decoding with CRC$16$ demonstrates a slight performance loss compared to the conventional SC and SRFSC decoders as a result of adding extra CRC bits. For polar codes of other lengths and rates, a similar trend can also be observed when comparing the BLER performance of different schemes.

\begin{figure}[t]
\centering
\scalebox{0.75}{\includegraphics{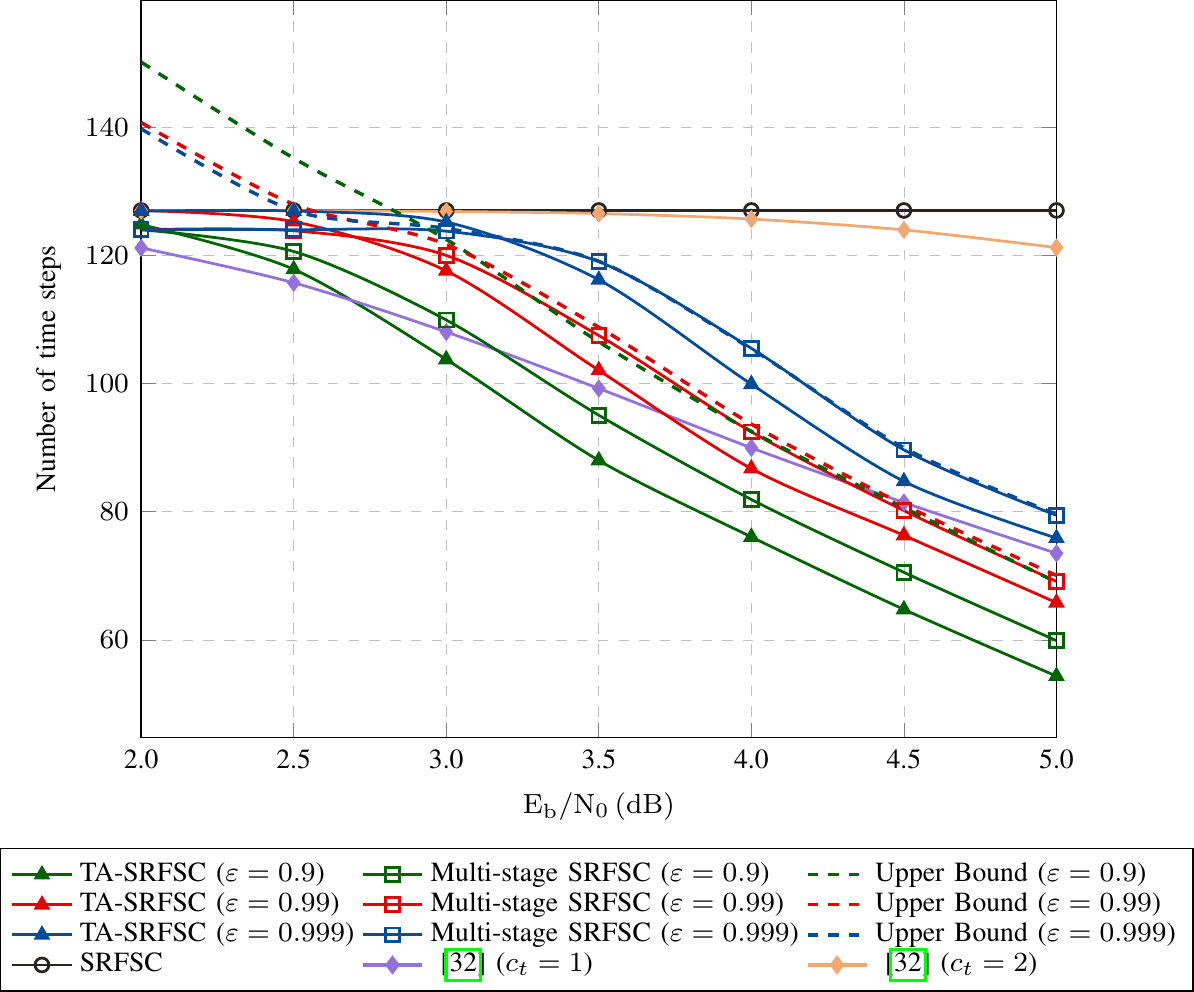}}
\setlength{\abovecaptionskip}{-5pt}
\caption{Average decoding latency of different decoding algorithms for the 5G polar code of length $N=1024$ and rate $R=1/2$.}
\setlength{\belowcaptionskip}{-5pt} 
\label{fig:TA-BS}
\vspace{-2em}
\end{figure}

Fig.~\ref{fig:TA-BS} presents the average decoding latency in terms of the required number of time steps for the proposed TA-SRFSC decoding and multi-stage decoding with CRC$16$. In particular, it compares them with the latency of SRFSC decoding and the decoder in \cite{li2018low} at different values of $\mathrm{E_b}/\mathrm{N_0}$ when $N=1024$ and $R=1/2$. It can be seen that the required number of time steps for the proposed TA-SRFSC decoding decreases as $\mathrm{E_b}/\mathrm{N_0}$ increases and is reduced by $40\%$ for $\varepsilon=0.999$, by $48\%$ for $\varepsilon=0.99$, and by $57\%$ for $\varepsilon=0.9$, compared to SRFSC decoding at $\mathrm{E_b}/\mathrm{N_0}=5$~dB. In addition, the proposed multi-stage SRFSC decoding reduces the required number of time steps by $37\%$ for $\varepsilon=0.999$, by $46\%$ for $\varepsilon=0.99$, and by $53\%$ for $\varepsilon=0.9$, with respect to SRFSC decoding at $\mathrm{E_b}/\mathrm{N_0}=5$~dB. The required number of time steps for the proposed multi-stage SRFSC decoding outperforms the method in \cite{li2018low} with $c_t=1$ by $19\%$ for $\varepsilon=0.9$ at $\mathrm{E_b}/\mathrm{N_0}=5$~dB while providing a significantly better BLER performance. Fig.~\ref{fig:TA-BS} also presents the approximate upper bound derived in (\ref{eq:latency2}). It can be seen in the figure that the upper bound in (\ref{eq:latency2}) becomes tighter as $\varepsilon$ increases.

\begin{figure}[t]
\centering
\scalebox{0.75}{\includegraphics{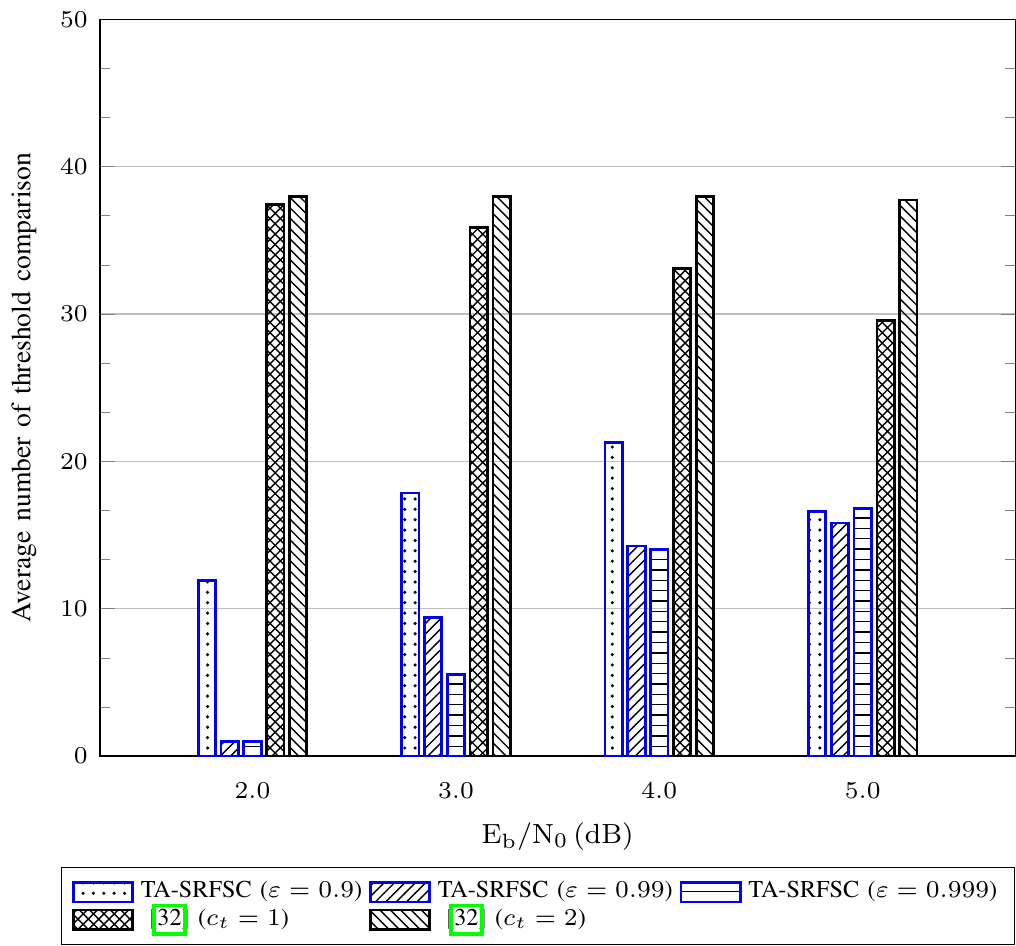}}
\caption{Average number of threshold comparisons of the proposed TA-SRFSC decoding in comparison with the hard-decision scheme in \cite{li2018low} for the 5G polar code of length $N=1024$ and rate $R=1/2$.}
  \label{fig:thresholdcomparison}
\end{figure}

Fig.~\ref{fig:thresholdcomparison} compares the average number of threshold comparisons in (\ref{eq:hard decision}) for the proposed TA-SRFSC decoder with $\varepsilon \in \{0.9,0.99,0.999\}$, and the decoder in \cite{li2018low} with $c_t\in\{1,2\}$ for the 5G polar code of length $N=1024$ and rate $R=1/2$. It can be seen that the proposed TA-SRFSC decoder shows significant benefits with respect to the decoder of \cite{li2018low} in terms of the average number of threshold comparisons. The TA-SRFSC decoder with $\varepsilon=0.9$ provides at least $36\%$ reduction with respect to \cite{li2018low} with $c_t=1$ while having a lower decoding latency. This means the decoder in \cite{li2018low} executes many unnecessary threshold comparison operations, while TA-SRFSC decoding only makes hard decisions when a node satisfies the condition in Equation (\ref{eq:ineq11}).

\section{Conclusion}
\label{sec:Conclu}

In this work, a new sequence repetition (SR) node is identified in the successive-cancellation (SC) decoding tree of polar codes and an SR node-based fast SC (SRFSC) decoder is proposed. In addition, to speed up the decoding of nodes with no specific structure, the SRFSC decoder is combined with a threshold-based hard-decision-aided (TA) scheme and a multi-stage decoding strategy. We show that this method further reduces the decoding latency at high SNRs. In particular, hardware implementation results on an FPGA for a polar code of length $1024$ with code rates $1/4$, $1/2$, and $3/4$ show that the proposed SRFSC decoder with merging branch operations requires up to $8\%$ fewer clock cycles and achieves $10\%$ higher maximum operating frequency compared to state-of-the-art decoders. In addition, the proposed TA-SRFSC decoding reduces the average decoding latency by $57\%$ with respect to SRFSC decoding at $\mathrm{E_b}/\mathrm {N_0}=5$~dB on a polar code of length $1024$ and rate $1/2$. This average latency saving is particularly important in real-time applications such as video. Future work includes the design of a fast SC list decoder using SR nodes.

\appendices

\section{Proof of Proposition~\ref{pr:SN}} \label{app:a}

Let $I_k$ denote the $k\times k$ identity matrix for $k\geq1$. Since the source node is the rightmost node in an SR node, the $g$ function calculation in (\ref{eq:g_function}) can be used as
{\small
    \begin{align*}
\alpha_r^E\left[1:2^r\right]=&\alpha_j^i\left[1:2^j\right]\times\left(I_{2^{j-1}}\otimes\left(\left(-1\right)^{\eta_{j-1}},1\right)^T\right)\\
&\times\left(I_{2^{j-2}}\otimes\left(\left(-1\right)^{\eta_{j-2}},1\right)^T\right)\times\cdots\\
&\times\left(I_{2^r}\otimes\left(\left(-1\right)^{\eta_r},1\right)^T\right).
    \end{align*}
}
Using the identity $\left(A\otimes B\right)\times\left(C\otimes D\right)=\left(A\times C\right)\otimes\left(B\times D\right)$ with $A=I_{2^{j-2}}$, $B=I_2\otimes\left(\left(-1\right)^{\eta_{j-1}},1\right)^T$, $C=I_{2^{j-2}}$, and $D=\left(\left(-1\right)^{\eta_{j-2}},1\right)^T$ results in
{\small
    \begin{align*}
\alpha_r^E\left[1:2^r\right]=&\alpha_j^i\left[1:2^j\right]\nonumber\times\cdots\\
&\Big[\!\left(I_{2^{j-2}}\times I_{2^{j-2}}\right)\otimes\left(I_2\otimes\left(\left(-1\right)^{\eta_{j-1}},1\right)^T\times\left(\left(-1\right)^{\eta_{j-2}},1\right)^T\right)\!\Big]\nonumber\\
&\times\left(I_{2^{j-3}}\otimes\left(\left(-1\right)^{\eta_{j-3}},1\right)^T\right)\times\cdots\times\left(I_{2^r}\otimes\left(\left(-1\right)^{\eta_r},1\right)^T\right),
    \end{align*}
}
which can be written as
{\small\begin{align*}
\alpha_r^E\left[1:2^r\right]=&\alpha_j^i\left[1:2^j\right]\times I_{2^{j-2}}\otimes\left(\left(\left(-1\right)^{\eta_{j-2}},1\right)^T\otimes\left(\left(-1\right)^{\eta_{j-1}},1\right)^T\right)\nonumber\\
&\times\left(I_{2^{j-3}}\otimes\left(\left(-1\right)^{\eta_{j-3}},1\right)^T\right)\times\cdots\times\left(I_{2^r}\otimes\left(\left(-1\right)^{\eta_r},1\right)^T\right),
    \end{align*}
}
where the identity $I_2\otimes\left(a_1,\dots,a_k\right)^T\times\left(b_1,b_2\right)^T=\left(b_1,b_2\right)^T\otimes\left(a_1,\dots,a_k\right)^T$ is used. Repeating the above procedures results in
{\small
\begin{align*}
\alpha_r^E\left[1:2^r\right]&=\alpha_j^i\left[1:2^j\right]\times I_{2^r}\!\otimes\!\left(\left(\left(-1\right)^{\eta_r},1\right)^T\!\otimes\!\cdots\otimes\left(\left(-1\right)^{\eta_{j-1}},1\right)^T\right)\nonumber\\
&=\alpha_j^i\left[1:2^j\right]\times \!I_{2^r}\!\otimes\!\left(\left(-1\right)^{{s}_l\left[1\right]}\!,\!\left(-1\right)^{{s}_l\left[2\right]}\!,\!\ldots\!,\!\left(-1\right)^{{s}_l\left[2^{j-r}\right]}\right)^{\!T}\!.
    \end{align*}
}
Thus for $k\in\left\{1,\dots,2^r\right\}$,
\begin{equation*}
\alpha_{r_l}^E\left[k\right]= \sum_{m = 1}^{2^{j-r}} \alpha_{j}^i\left[\left(k-1\right)2^{j-r}+m\right] \left(-1\right)^{s_l[m]}.
\end{equation*}

\section{Explanation of Observation~\ref{pr:LB}} \label{app:b}

To explain Observation~\ref{pr:LB}, a lemma is first introduced as follows.

\noindent
\textbf{Lemma 1.}
\textit{Under the Gaussian approximation assumption in \eqref{eq:gaussian}, for any node $\mathcal{N}_j^i$ whose $2^j$ bits undergo a hard decision in (\ref{eq:hard decision}), assuming all prior bits are decoded correctly, the probability of correct decoding can be calculated as $\left(\frac{\widetilde P_c}{\widetilde P_e+\widetilde P_c}\right)^{2^j}$.}

\begin{IEEEproof}
In accordance with Fig.~\ref{fig:distribution}, for any node $\mathcal{N}_j^i$, considering all the previous bits are decoded correctly, the probability that the $k$-th bit ($1\leq k\leq 2^j$) in the node undergoes a hard decision is $\widetilde P_c+\widetilde P_e$. Moreover, The probability of a correct hard decision for the $k$-th bit in the node is $\widetilde P_c$, regardless of the value of $\beta_j^i\left[k\right]$. Thus, the conditional probability that a hard decision on the $k$-th bit is correct given that the $k$-th bit undergoes a hard decision is $\frac{\widetilde P_c}{\widetilde P_e+\widetilde P_c}$. Since the LLR values of bits in a node are independent of each other, the conditional probability that hard decisions on all the $2^j$ bits of node $\mathcal{N}_j^i$ are correct given that all its $2^j$ bits undergo hard decisions can be calculated as $\left(\frac{\widetilde P_c}{\widetilde P_e+\widetilde P_c}\right)^{2^j}$.
\end{IEEEproof}
To have a probability of correct decoding of at least $\varepsilon$ for all the nodes that undergo hard decision in a polar code of length $2^n$, any such node $\mathcal{N}_j^i$ is assumed to have the probability of correct decoding of at least $\sqrt[2^{\left(n-j\right)}]\varepsilon$. 
Therefore and by using the result in Lemma~1, we have
\begin{equation} \label{eq:ineq1sim}
\left(\frac{\widetilde P_c}{\widetilde P_e+\widetilde P_c}\right)^{2^j}\geq\sqrt[2^{\left(n-j\right)}]\varepsilon,
\end{equation}
which is equivalent to
\begin{equation} \label{eq:ineq1} \frac{\widetilde P_c}{\widetilde P_e+\widetilde P_c}\geq\sqrt[2^{n}]\varepsilon.
\end{equation}
If $m_j^i\leq2c^2$, then $T=-m_j^i+c\sqrt{2m_j^i}$, $\widetilde P_c=Q\left(c-\sqrt{2m_j^i}\right)$, and $\widetilde P_e=Q\left(c\right)$. When $0.5<\varepsilon<1$, (\ref{eq:ineq1}) can be written as
\begin{equation} \label{eq:ineq4}
\frac{1}{2}\left[c-Q^{-1}\left(\frac{Q\left(c\right)}{\frac1{\sqrt[2^n]{\varepsilon}}-1}\right)\right]^2\leq m_j^i\leq2c^2,
\end{equation}   
which requires
\begin{equation}     \label{eq:ineq66}
Q\left(c\right)\leq1-\sqrt[2^n]{\varepsilon}.
\end{equation}
If $m_j^i\geq2c^2$, then $T=m_j^i-c\sqrt{2m_j^i}$, $\widetilde P_c=Q\left(-c\right)$, and $\widetilde P_e=Q\left(\sqrt{2m_j^i}-c\right)$. Thus (\ref{eq:ineq1}) can be written as
\begin{equation} \label{eq:ineq8}
m_j^i\geq\max\left\{2c^2,\frac12\left[c+Q^{-1}\left(\left(\frac1{\sqrt[2^n]{\varepsilon}}-1\right)Q\left(-c\right)\right)\right]^2\right\}.
\end{equation}
If (\ref{eq:ineq66}) holds and by using the fact that $Q(-c) = 1 - Q(c)$, then
\begin{equation} \label{eq:ineq9}
2c^2\geq\frac12\left[c+Q^{-1}\left(\left(\frac1{\sqrt[2^n]{\varepsilon}}-1\right)Q\left(-c\right)\right)\right]^2.
\end{equation} 
Thus $m_j^i\geq2c^2$, which always holds based on the initial assumption. Therefore, it suffices that
\begin{equation} \label{eq:ineq111}
m_j^i\geq\frac12\left[c-Q^{-1}\left(\frac{Q\left(c\right)}{\frac1{\sqrt[2^n]{\varepsilon}}-1}\right)\right]^2,
\end{equation}
and (\ref{eq:ineq66}) to ensure (\ref{eq:ineq1}). In other words, assuming all previous bits are decoded correctly, the probability that any node $\mathcal{N}_j^i$ that undergo hard decision (\ref{eq:hard decision}) in the decoding process is decoded correctly is lower bounded by $\sqrt[2^{\left(n-j\right)}]\varepsilon$ if (\ref{eq:ineq66}) and (\ref{eq:ineq111}) are satisfied.

\section{Explanation of Observation~\ref{pr:BLER}} \label{app:c}

Note that based on Observation~\ref{pr:LB}, any node $\mathcal{N}_j^i$ that is decoded using (\ref{eq:hard decision}) has a probability of correct hard decision of approximately greater than or equal to $\sqrt[2^{\left(n-j\right)}]\varepsilon$. For any node that undergoes the SRFSC decoding, the probability of correct decoding is determined by the error rate of SRFSC decoding. Thus, the probability of correct decoding for TA-SRFSC decoding is approximately greater than or equal to $\varepsilon\left(1-{\mathrm{BLER}}_{\mathrm{SRFSC}}\right)$. Consequently, ${\mathrm{BLER}}_{\mathrm{TA-SRFSC}}\lesssim 1-\varepsilon\left(1-{\mathrm{BLER}}_{\mathrm{SRFSC}}\right)$.

\ifCLASSOPTIONcaptionsoff
  \newpage
\fi

\bibliography{myreference}

\end{document}